\documentclass[12pt,a4paper,final]{iopart}

\usepackage{iopams}  
\usepackage{float}  
\usepackage{graphicx}
\usepackage[breaklinks=true,colorlinks=true,linkcolor=blue,urlcolor=blue,citecolor=blue]{hyperref}
\usepackage{color}
\usepackage{url}

\begin{document}

\title[Microtearing Thresholds and Second-Stable Ballooning in the DIII-D Pedestal: Reduced Modeling and Core-Edge Implications]{Microtearing Thresholds and Second-Stable Ballooning in the DIII-D Pedestal: Reduced Modeling and Core-Edge Implications}

\author{D. R. Hatch$^{1*,2}$}
\author{L. A. Leppin$^{1,3}$}
\author{M. T. Kotschenreuther$^{2}$}
\author{S. Houshmandyar$^{1}$}
\author{S. M. Mahajan$^{1,2}$}
\author{J. Schmidt$^{1}$}
\author{P.-Y. Li$^{1}$}

\address{$^1$Institute for Fusion Studies, University of Texas at Austin, Austin, Texas, USA}
\address{$^2$ExoFusion, Austin, Texas, USA}
\address{$^3$Oden Institute, University of Texas at Austin, Austin, Texas, USA}
\ead{$^*$drhatch@austin.utexas.edu}

\begin{abstract}
Global and local linear gyrokinetic simulations of 42 pedestal equilibria from three DIII-D discharges are used to investigate pedestal stability and its impact on pedestal structure and confinement. Microtearing modes (MTMs) and kinetic ballooning modes (KBMs) represent the main ion scale instabilities. For all three discharges, MTMs lie near a stability boundary in the mid-pedestal and exhibit threshold behavior, with growth rates increasing at and beyond pre-ELM pressure gradients. Pedestal MTMs retain conventional signatures but also show enhanced particle transport and partial density-gradient drive, indicating they can constrain pedestal {\it pressure} rather than electron temperature alone. KBMs are typically second-stable in this region due to low magnetic shear and large pressure gradients, though they can become active near the pedestal foot where magnetic shear is higher. These findings suggest MTMs play the role of inter-ELM pressure limit in the mid-pedestal when KBM is second stable. A preliminary quasilinear mixing-length transport model, with properly tuned free parameters, reproduces experimental temperature and density profiles when coupled to ASTRA. When applied to a case with doubled separatrix density, the model predicts reduced pedestal pressure consistent with ITPA H-mode confinement trends, attributable to increased MTM and ETG transport. These results clarify pedestal-limiting mechanisms and establish a physics-based link between separatrix conditions, pedestal structure, and global confinement.  This work lays the foundation for new predictive modeling capabilities for core-edge integration in burning plasma regimes.   
          
\end{abstract}

\vspace{2pc}
\section{Introduction} \label{sec:introduction}


This work investigates the main gyrokinetic instabilities and their role in the pedestal of three DIII-D discharges for which previous studies~\cite{hassan_NF_21,kotschenreuther_19,halfmoon_pop_22} have found close connections between microtearing modes and magnetic fluctuation data.  The gyrokinetic code, \textsc{GENE}~\cite{gene_ref1,goerler_11}, is used to study both global and local instabilities.  

The main innovations in methodology are as follows.  
First, instead of investigating a single equilibrium fit in the pre-ELM phase of the discharge, an ensemble of equilibria is constructed in the vicinity of the pre-ELM state, spanning combinations of both higher and lower temperature and density gradients.  These equilibria are used to investigate the behavior of instabilities approaching and surpassing the saturated phase of the inter-ELM pedestal evolution.  Second, the resulting linear gyrokinetic simulations are used to formulate surrogate models for mixing length diffusivities, which are then coupled to a transport code, ASTRA~\cite{ASTRA,fable_2013,luda_20,Tardini2026}.  This enables dynamic modeling of the pedestal to probe the effects of these instabilities on pedestal structure, confinement, and core-edge integration.  

The gyrokinetic simulations identify two main instabilities: microtearing modes (MTM) and kinetic ballooning modes (KBM).  There is substantial material in the literature describing the role of these instabilities in the pedestal.  The KBM has been proposed as the main inter-ELM pressure constraint and is a central feature of EPED-like models~\cite{snyder_09,snyder_09b} and other analyses of pedestal stability~\cite{hatch_15,Parisi_2024a,Parisi_2024b,saarelma_13,saarelma_17,tzanis_25}.  Several papers have made concrete connections between MTM properties and experimentally-observed magnetic fluctuations~\cite{ perez_03,diallo_15,laggner_16,laggner_19} in the pedestal~\cite{hatch_16,kotschenreuther_19,chen_19,chen_20,hatch_21,larakers_21,hassan_NF_21,halfmoon_pop_22,curie_22,curie_RIP_22,dominski2020identification}. Typically MTM has been proposed as a constraint on the electron temperature in the pedestal.  This paper builds on this body of work and revises it on some important points.  We find that MTMs are capable of acting as a constraint on pedestal {\it pressure}, not just pedestal temperature.  This occurs in the mid-pedestal region where KBMs are often in a second stability regime.  The KBM is active primarily at the foot of the pedestal where magnetic shear is high near the separatrix.  

We verify precise criteria for discriminating between the two instabilities by using a clustering algorithm in comparison with a fundamental understanding of the underlying physics. Both local and global simulations produce the same general picture for MTMs: they become unstable for equilibria at or near the pre-ELM pedestal with mixing length diffusivities increasing sharply with pressure gradients---in other words, MTM exhibits threshold behavior that corresponds to the experimental dynamics.  For KBM, the picture is much less clear.  Global simulations do not identify strongly unstable KBM.  Local simulations identify strongly unstable KBM at weaker pressure gradients, but they stabilize as the pedestal enters second stability in the saturated pre-ELM phase.  KBM exhibits threshold behavior only near the pedestal foot where magnetic shear remains high.  

The implications for pedestal structure are investigated by coupling quasilinear mixing length diffusivities with ASTRA for one of the DIII-D discharges.  Neoclassical transport and a model for pedestal ETG transport are also included in the modeling.  The recycling particle source is estimated from interpretive SOLPS modeling~\cite{TPT,guttenfelder_NF_21}.  For reasonable choices of free parameters, the model closely reproduces both the electron temperature and density profiles of the pre-ELM pedestal.  As a further physics test, the same model is applied to a scenario with higher separatrix density, which results in degraded confinement at a level qualitatively consistent with experimental observations.   

The major conclusions of this paper are as follows: for three DIII-D discharges, clear threshold behavior is identified for MTM near the pre-ELM phase of the discharge.  Such behavior is not identified for KBM in the steep gradient region.  Rather, KBM appears to be most active at the foot of the pedestal where magnetic shear remains high.  The properties of the MTM are somewhat distinct from previous analysis.  It produces relatively more particle transport than expected and also is driven by {\it pressure} gradients, not only electron temperature gradients. Consequently, it is adequate to constrain the inter-ELM pressure profile across much of the pedestal by itself without appealing to an additional instability like KBM.  The properties of MTM result in higher transport and reduced confinement as separatrix density increases, which may offer an important physics component explaining issues of core-edge integration.        
This paper is outlined as follows:  the experimental scenarios are described in Sec.~\ref{sec:experimental}.  Linear gyrokinetic analysis is described in Sec.~\ref{sec:linear}.  A quasilinear mixing length transport model is described in Sec.~\ref{sec:Dmix}.  Profile predictions using ASTRA are described in Sec.~\ref{sec:profile_predictions}.  Some broader discussion is provided in Sec.~\ref{sec:discussion} followed by a summary and conclusions in Sec.~\ref{sec:summary}.

\section{Description of Experimental Discharges}
\label{sec:experimental}

This work focuses on three DIII-D discharges: 162940, 174082, and 153764.  Discharge 174082 was the reference shot for a fueling study~\cite{nelson_NF_20}.  Discharge 162940 was part of an experiment designed to probe the effect of divertor closure~\cite{moser_pop_20}.  It has also been the focus of several gyrokinetic pedestal studies~\cite{guttenfelder_NF_21,hassan_NF_21,hassan_pop_21}. Discharge 153764 was the subject of an extensive study of magnetic fluctuations~\cite{diallo_15}.  All three discharges were analyzed with both gyrokinetic simulations and interpretive edge modeling as part of the 2019 US-DOE theory performance target~\cite{TPT} and clear connections were identified between gyrokinetic simulations and signatures in the magnetic fluctuations~\cite{hassan_NF_21,halfmoon_pop_22,kotschenreuther_19}.  

\begin{figure}[H]
    \centering
    \includegraphics[scale=0.8]{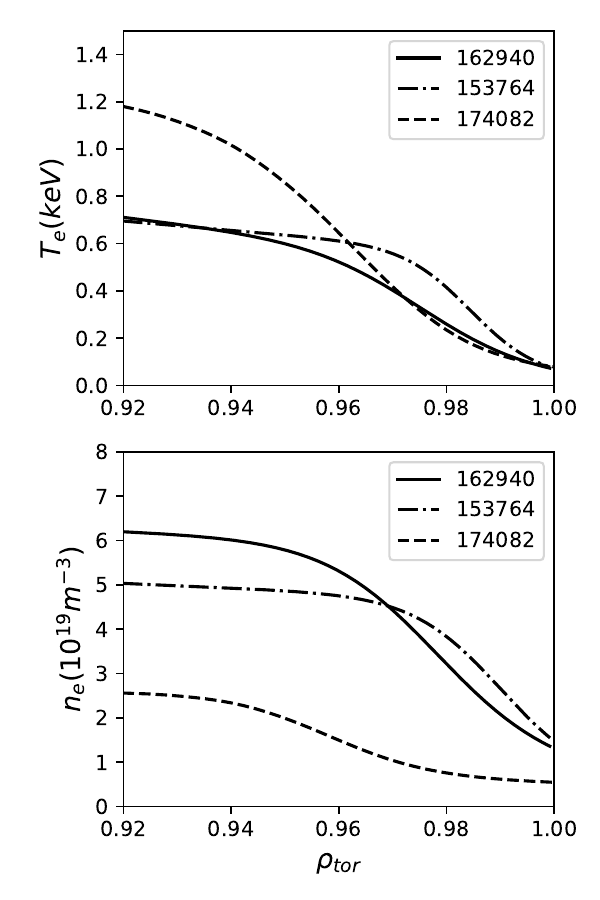}
    \caption{\label{profiles_all}  Pre-ELM electron temperature (top) and density (bottom) profiles for the three DIII-D discharges.    }
\end{figure}

\begin{figure}[H]
    \centering
    \includegraphics[scale=0.8]{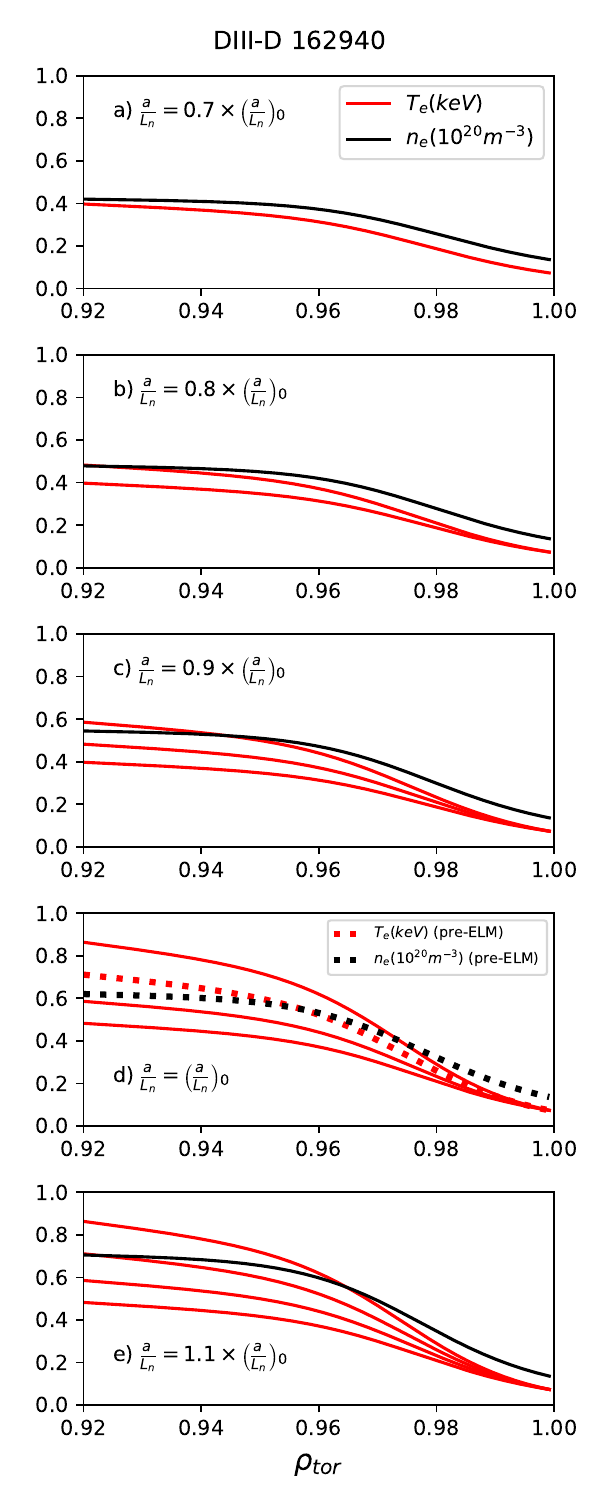}
    \caption{\label{profiles_ensemble}  The ensemble of density and temperature profiles for DIII-D shot 162940.  Each panel shows a single density profile along with all corresponding temperature profiles.  These are parameterized by the variations shown in Table~\ref{tab:variations}.    }
\end{figure}

Pre-ELM electron temperature and density profiles for the three discharges are shown in Fig.~\ref{profiles_all}.  In order to probe instability behavior in a space of nearby equilibria, an ensemble of fully self-consistent equilibria was generated for each discharge.  The electron density and temperature profiles were varied by modifying the gradient scale lengths by a constant factor across the pedestal: 

\begin{eqnarray}
 \omega_{Te} &=& \alpha_{T} \omega_{Te0} \\
 \omega_{ne} &=& \alpha_n \omega_{ne0}
\label{profile_mod}
\end{eqnarray}
where $\omega_{Te0} = -\frac{d \mathrm{ln} T_{e0}}{d \rho_t}$, $T_{e0}$ is experimental pre-ELM profile fit, the radial coordinate $\rho_t$ is the square root of the normalize toroidal magnetic flux.  By this method, a set of 14 equilibria were generated for each discharge as summarized in Table~\ref{tab:variations}.  The profile variations are shown in Fig.~\ref{profiles_ensemble} for discharge 162940.  The same algorithm was applied for the other discharges (not shown).  

\begin{table}[h]
\centering
\begin{tabular}{|c|c|c|c|c|c|c|c|c|c|c|c|c|c|c|}
\hline
$\alpha_n$ & 0.7 & 0.8 & 0.8 & 0.9 & 0.9 & 0.9 & 1.0 & 1.0 & 1.0 & 1.0 & 1.1 & 1.1 & 1.1 & 1.1 \\
\hline
$\alpha_T$ & 0.7 & 0.7 & 0.8 & 0.7 & 0.8 & 0.9 & 0.8 & 0.9 & 1.0 & 1.1 & 0.8 & 0.9 & 1.0 & 1.1 \\
\hline
\end{tabular}
\label{tab:variations}
\caption{Variations of pedestal density and temperature gradients applied to all three experimental scenarios.}
\end{table}

\section{Linear Gyrokinetic Simulations}
\label{sec:linear}

This section describes linear gyrokinetic simulations of the ensembles of equilibria for the three discharges.  Numerical details of the simulations are described in ~\ref{appendix:numerical}.  In order to define quantitative criteria for mode classification, k-means clustering is used to guide physics-based mode identification criteria as described in Sec.~\ref{sec:clustering}.  Following that, Sections~\ref{sec:global_linear} and~\ref{sec:local_linear} describe the resulting instabilities and their behavior throughout the space of equilibria described above.  

\subsection{Clustering}
\label{sec:clustering}

In this subsection, k-means clustering is applied to the local linear simulations to both guide and verify physics-based criteria for mode classification.  This analysis considers only ion scales $k_y \rho_s \approx 0.01-1$.  Smaller scale instabilities are invariably some kind of ETG mode, which have been studied extensively for this and other discharges elsewhere~\cite{jenko_09,hatch_15,guttenfelder_NF_21,hassan_pop_21,parisi_NF_20,hatch_22,hatch_24}.  The following parameters were used for the classification analysis: $\hat{Q}_{EM} = Q_{eEM} / (Q_{eES} + Q_{iES})$, parity of the magnetic vector potential, $P(A_{||}) = \int dz A_{||} / \int dz |A_{||}|$, parity of the electrostatic potential, $P(\phi) = \int dz \phi / \int dz |\phi|$, a normalized frequency $\omega / \omega_{*e}$, the parallel electric field $\hat{E}_{||} = \frac{\int dz | -\partial_z phi + i \omega A_{||}|}{\int dz | \partial_z phi| + \int dz |i \omega A_{||}|}$, and the diffusivity ratio $D_e/\chi_e$.  

Ref.~\cite{kotschenreuther_19} defines several `fingerprints' for distinguishing between various instabilities.  MTM is noted for high electromagnetic heat flux ($\hat{Q}_{EM}$), frequencies in the vicinity of $\omega_{*e} (a/c_s)= k_y \rho_s  (\omega_{ne}+\omega_{Te})$, low particle diffusivity ($D_e / \chi_e \ll 1$), and at least partial tearing parity ($P(A_{||}) > 0$).  KBM is characterized by small parallel electric field $\hat{E}_{||} \ll 1$, comparable particle and thermal diffusivities ($D_e/\chi_e \sim 1$), and strong ballooning parity.  The present analysis largely aligns with these criteria but adds some nuance and quantifies effective boundaries as described below.

\begin{figure}[H]
    \centering
    \includegraphics[scale=0.8]{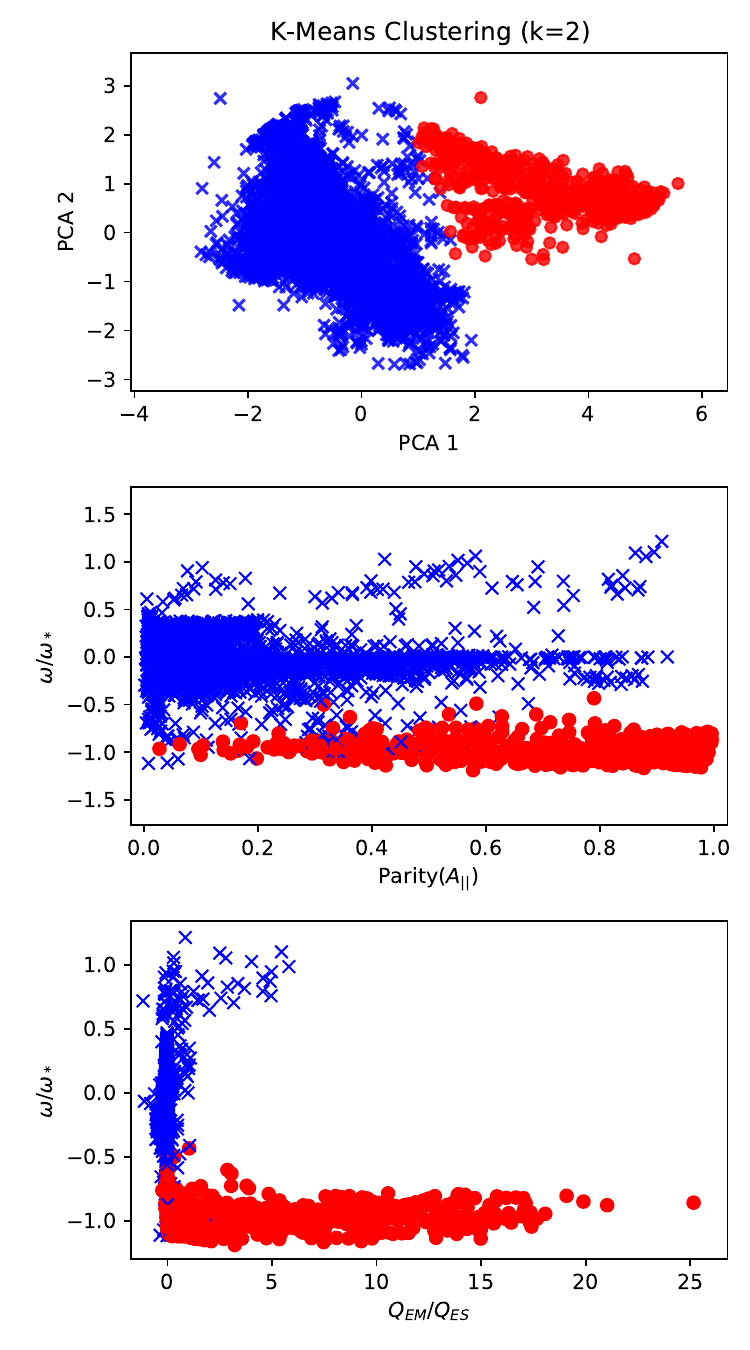}
    \caption{\label{clustering}  Plots related to classification and k-means clustering.  Red is MTM and blue is all other instabilities (notably, KBM).  Top: two-dimensional projection of the six-dimensional feature space onto the first two principal components (PCA), with points colored by k-means cluster assignment (k = 2). Middle: plot of the normalized frequency and parity of $A_{||}$ with symbols and colors denoting the two clusters.  Bottom: plot of the normalized frequency and electromagnetic heat flux ratios with symbols and colors denoting the two clusters.     }
\end{figure}

A k-means clustering algorithm was applied to the data for $k=2-9$ clusters to explore the number of distinct categories that best classifies the data.  The clear winner is $k=2$ (silhouette = 0.495 while $k>2$ produces silhouette values less than 0.4).  Results for $k=2$ are shown in Fig.~\ref{clustering} where distinct regions can be identified, shown in red and blue.  In order to highlight some of the main physical parameter dependencies, the lower two plots show the normalized frequency ($\omega / \omega_*$) plotted against the mode parity (middle) and electromagnetic flux (bottom).  Cluster means are shown in Table~\ref{tab:cluster_means}, demonstrating alignment with physics-based expectations for MTM and KBM.  

Based on this analysis, we define the following criteria to identify MTM: 
\begin{eqnarray}
\hat{Q}_{EM} &>& 0.2 \label{QEM_criterion} \\
\omega/\omega_{*e} &<& -0.5 \label{omega_criterion} \\
P(A_{||}) &>& 0.15 \label{parity_criterion}
\end{eqnarray}
These criteria result in the following comparisons with the clustering algorithm: the physics criteria and the clustering algorithm agree in 5054 out of 5154 simulations.  In 587 simulations, they agree in classifying the instabilities as MTM.  In 4467 simulations they agree in classifying the instabilities as not MTM.  There is disagreement in 100 of the simulations (7 of which the physics criteria indicate MTM and 93 of which the physics criteria indicate the instabilities as not MTM).      





Although the algorithm identifies two categories as the cleanest categorization, a three category analysis is also informative.  It results in a subdivision of the non-MTM modes into a more electromagnetic category (i.e., more KBM-like) and a more electrostatic category, consistent with physics intuition.  Cluster means are shown in Table~\ref{tab:cluster_means_k3}, where the differences between cluster 0 and 2 in $\hat{E_{||}}$ and $D_e/\chi_e$ are notable.  

\begin{table}[htbp]
\centering
\caption{Cluster means for k-means clustering with two categories.}
\begin{tabular}{lrrrrrr}
\hline
\textbf{Cluster} & $\omega_{\mathrm{norm}}$ & $\mathrm{parity}_{A_\parallel}$ & $\mathrm{parity}_{\phi}$ & $\log Q_{EM}$ & $\hat{E}_{\parallel}$ & $D_e/\chi_e$ \\
\hline
0 & -0.0376 & 0.1431 & 0.7074 & -3.8950 & 0.6596 & 0.5439 \\
1 & -1.0023 & 0.7453 & 0.3605 & 0.9915 & 0.4300 & 0.0873 \\
\hline
\end{tabular}
\label{tab:cluster_means}
\end{table}

\begin{table}[htbp]
\centering
\caption{Cluster means for k-means clustering with three categories.}
\begin{tabular}{lrrrrrr}
\hline
\textbf{Cluster} & $\omega_{\mathrm{norm}}$ & $\mathrm{parity}_{A_\parallel}$ & $\mathrm{parity}_{\phi}$ & $\log Q_{EM}$ & $\hat{E}_{\parallel}$ & $D_e/\chi_e$ \\
\hline
0 & -0.0317 & 0.0771 & 0.8099 & -2.9259 & 0.5384 & 0.7183 \\
1 & -1.0028 & 0.7225 & 0.3497 & 1.0887  & 0.4373 & 0.0860 \\
2 & -0.0489 & 0.2646 & 0.5182 & -5.6767 & 0.8825 & 0.2222 \\
\hline
\end{tabular}
\label{tab:cluster_means_k3}
\end{table}

\subsection{Growth Rates and Frequencies}
\label{sec:growth_rates}

\subsubsection{Global Linear}
\label{sec:global_linear}

\begin{figure}[H]
    \centering
    \includegraphics[scale=0.8]{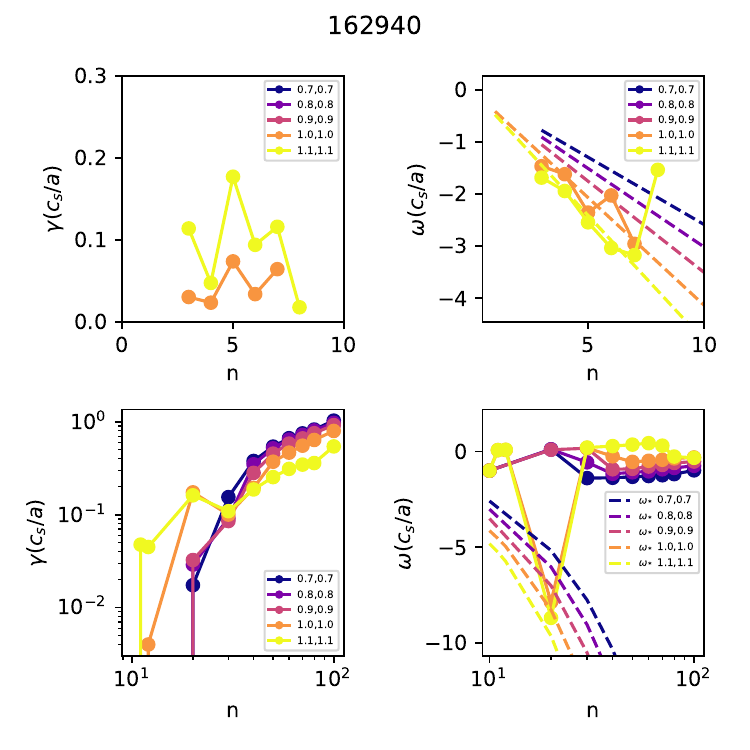}
    \caption{\label{global162940} Global growth rates (left) and frequencies (right) for low toroidal mode numbers (top) and high toroidal mode numbers (bottom) for \textsc{GENE} simulations of DIII-D discharge 162940.  Colors denote the pressure gradient variations with lighter colors (yellow) corresponding to steeper pressure gradients.  Note that parameter points with negative growth rates (i.e., no instabilities) are not plotted.  Compare with mixing length estimates shown in Fig.~\ref{global_Dmix_162940}, which will be discussed later.   }
\end{figure}

\begin{figure}[H]
    \centering
    \includegraphics[scale=0.8]{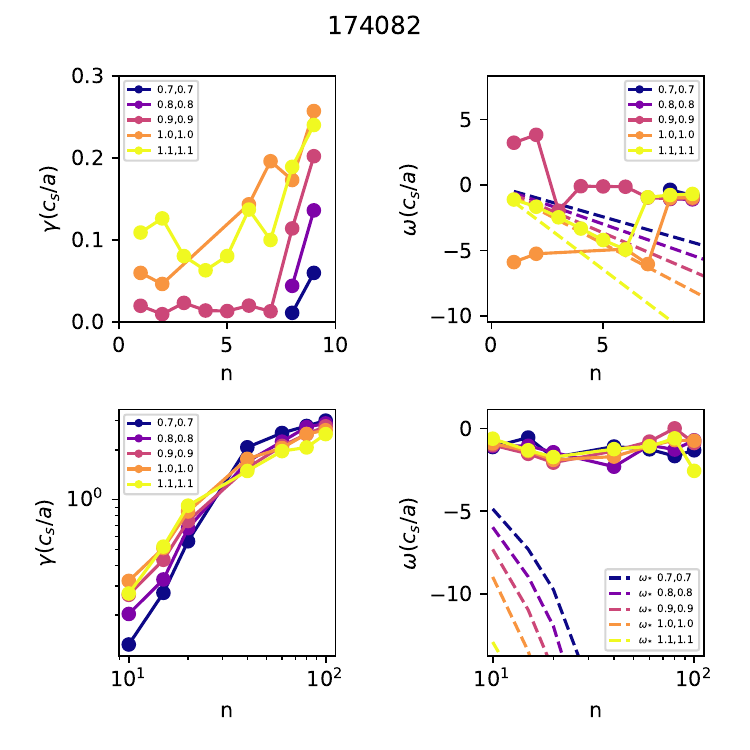}
    \caption{\label{global174082} Global growth rates (left) and frequencies (right) for low toroidal mode numbers (top) and high toroidal mode numbers (bottom) for \textsc{GENE} simulations of DIII-D discharge 174082.  Colors denote the pressure gradient variations with lighter colors (yellow) corresponding to steeper pressure gradients.  Note that parameter points with negative growth rates (i.e., no instabilities) are not plotted.  Compare with mixing length estimates shown in Fig.~\ref{global_Dmix_174082}, which will be discussed later.   }
\end{figure}

\begin{figure}[H]
    \centering
    \includegraphics[scale=0.8]{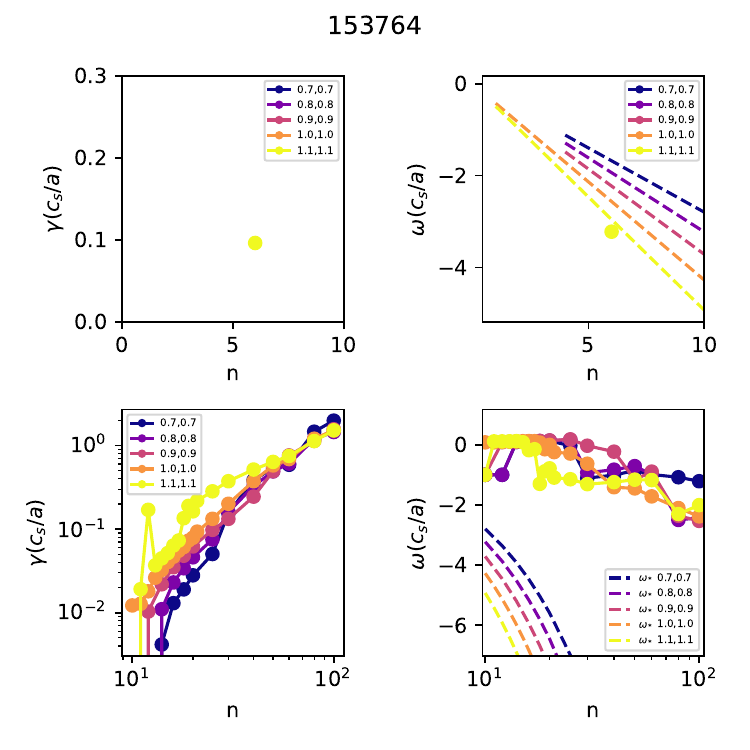}
    \caption{\label{global153764} Global growth rates (left) and frequencies (right) for low toroidal mode numbers (top) and high toroidal mode numbers (bottom) for \textsc{GENE} simulations of DIII-D discharge 153764.  Colors denote the pressure gradient variations with lighter colors (yellow) corresponding to steeper pressure gradients.  Note that parameter points with negative growth rates (i.e., no instabilities) are not plotted.  Compare with mixing length estimates shown in Fig.~\ref{global_Dmix_153764}, which will be discussed later.   }
\end{figure}

This section shows growth rates and frequencies for selected equilibria.  We begin with global growth rates and frequencies, which are shown in Figs.~\ref{global162940},~\ref{global174082},~\ref{global153764}.  In these figures, lower toroidal mode numbers ($n\sim 1-10$) are shown in the top row and higher toroidal mode numbers ($n \sim 10-100$) are shown in the bottom row.  Although growth rates are smaller in the low-n range, these modes dominate the mixing-length transport estimates as will be described in Sec.~\ref{sec:Dmix} (this is also consistent with nonlinear simulations).The corresponding equilibria are denoted by color, with lighter (yellow) colors denoting higher pressure gradients.  Stable simulations are not plotted.  MTM can be identified by their frequencies, which are close to $\omega_{*e}$ (indicated by dashed lines).  In the $n \sim 1-10$ range, MTMs become unstable at or near the pre-ELM equilibrium for all cases.  MTMs become unstable for several toroidal mode numbers for the two steepest gradient cases.  For 153764, there is only a single unstable mode identified in this low-$n$ range over the entire set of equilibria and this mode is a MTM.  In short: for all three discharges, the global stability analysis identifies low-$n$ MTMs becoming unstable at or just beyond the pre-ELM state, {\it suggesting that the MTMs are responsible for saturating the inter-ELM $T_e$ and possible $P_e$ profiles}.  This is consistent with experimental observations.  Ref.~\cite{diallo_15} shows correlation between the saturation of the inter-ELM gradients and onset of magnetic fluctuations, which were later identified as MTM~\cite{kotschenreuther_19}.  Likewise, Ref.~\cite{chen_20} connects internal magnetic fluctuation measurements with properties of MTM and demonstrates the effect of these fluctuations on confinement.  

Low-$n$ KBM instabilities are not identified in these global simulations.  This may be because (1) KBM are actually stable for these scenarios or (2) it may be due to the limitations of the global simulations.  Regarding the latter, the simulations enforce Dirichlet boundary conditions at the ends of the domain, which could stabilize MHD-like modes which are sensitive to the vacuum solution extending to the wall.  However, since KBM instabilities are radially broad and the sharp pressure gradients are localized in a narrow pedestal region, one could argue that the low-$n$ KBMs are indeed stable and speculate that if such a mode became unstable, it would manifest itself as an ELM rather than an inter-ELM constraint.  We will discuss the possible role of KBM in more depth below, but for now we make the unambiguous observation that MTM exhibits threshold behavior for all three discharges.        

\subsubsection{Local Linear}
\label{sec:local_linear}

\begin{figure}[H]
    \centering
    \includegraphics[scale=0.8]{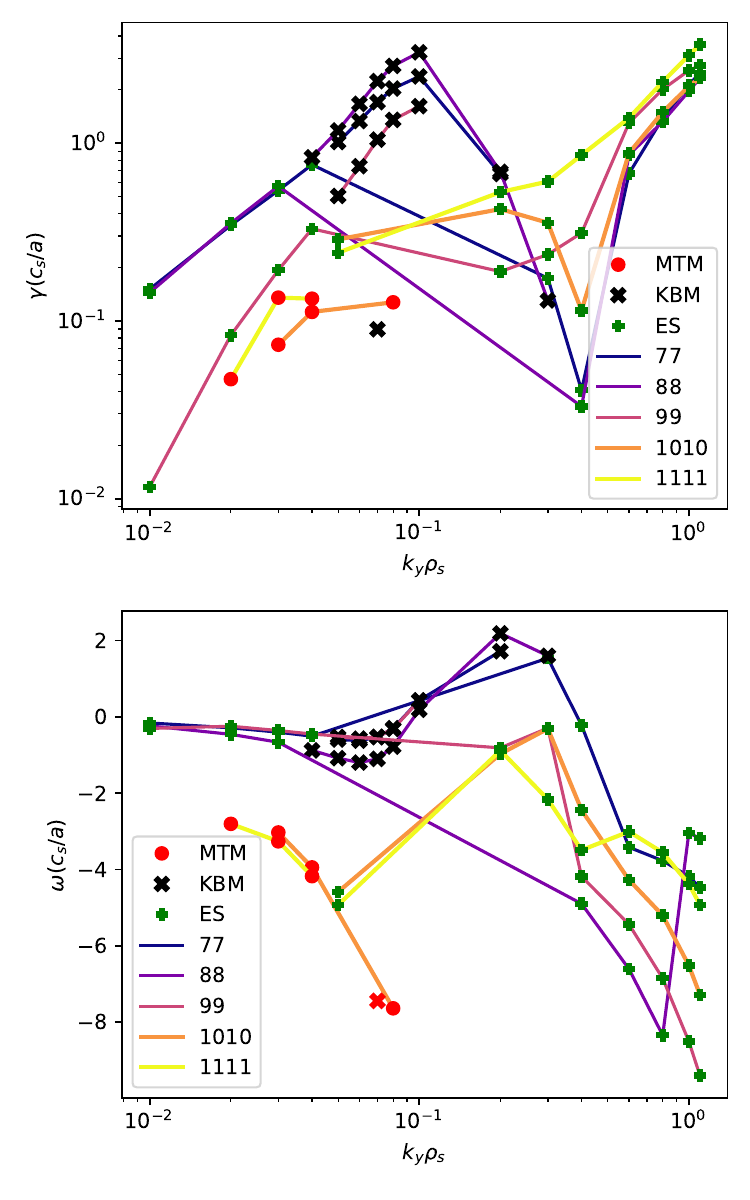}
    \caption{\label{local162940} Growth rates (top) and frequencies (bottom) for local linear \textsc{GENE} simulations for DIII-D shot 162940.  Symbols denote instability type while line color denotes pressure gradient variations (steeper gradients in yellow).  This is for the radial position $\rho_{tor} = 0.98$.   }
\end{figure}

Given the extensive dataset of local linear simulations (spanning 16 $k_y$ wavenumbers over 14 equilibria for each of three discharges) we restrict ourselves to a single radial location to illustrate some key points.  The mixing length diffusivities shown in Sec.~\ref{sec:Dmix} will provide a more comprehensive view.  Fig.~\ref{local162940} shows growth rates (top) and frequencies (bottom) for 162940 at $\rho_{tor} = 0.98$ over a large range of wavenumbers for five equilibria with increasing pressure gradients (ranging from $\alpha_T = \alpha_n = 0.7$ to $\alpha_T = \alpha_n = 1.1$).  For the ion temperature profile, the same $\alpha_T$ is applied so that its gradient variations track those of the electron temperature profile.  In this figure, MTMs are denoted in red, KBM in black, and electrostatic (ES) modes in green (the ES modes are defined as non-MTMs with $\hat{E}_{||} > 0.2$).   As with many earlier plots, the line colors correspond to the equilibrium, with lighter (yellow) lines denoting steeper pressure gradients.  In contrast with the global analysis, KBM are prominent in the instability spectrum.  However, they are only unstable at {\it weak} pressure gradients and become stabilized as the pedestal enters second stability at steeper pressure gradients (the second stability picture will be examined in more depth in the next section).  

As may be expected from the extreme pedestal gradients and narrow pedestal region, the flux tube approximation has limitations when applied in this region.  This is particularly notable for KBM instabilities, which are strongly ballooning at the outboard mid-plane and are characterized by broad radial eigenmodes.  Often, the eigenmodes identified in flux tube simulations have radial widths that are broader than the pedestal itself, in which case, one might be skeptical of the physical validity of the instability.  This is discussed in some depth in Ref.~\cite{hatch_21}, where extremely unstable MHD-like modes are identifies in local simulations but not global simulations.  These extremely unstable modes are not consistent with the underlying physics scenario (a low-transport pedestal) and were dismissed on this basis, justified by their extremely broad eigenmodes.  This likely the main reason why KBM is more prominent in the flux tube simulations than the global simulations.  

In contrast with the dramatically different manifestations of KBM in global and local simulations, the picture for MTM is basically consistent: MTM becomes unstable for the pre-ELM equilibrium and becomes more virulent as the pressure gradient increases---i.e., it exhibits threshold behavior.   

\subsection{MTM and KBM in Second Stability}
\label{sec:s_alpha}

\begin{figure}[H]
    \centering
    \includegraphics[scale=0.8]{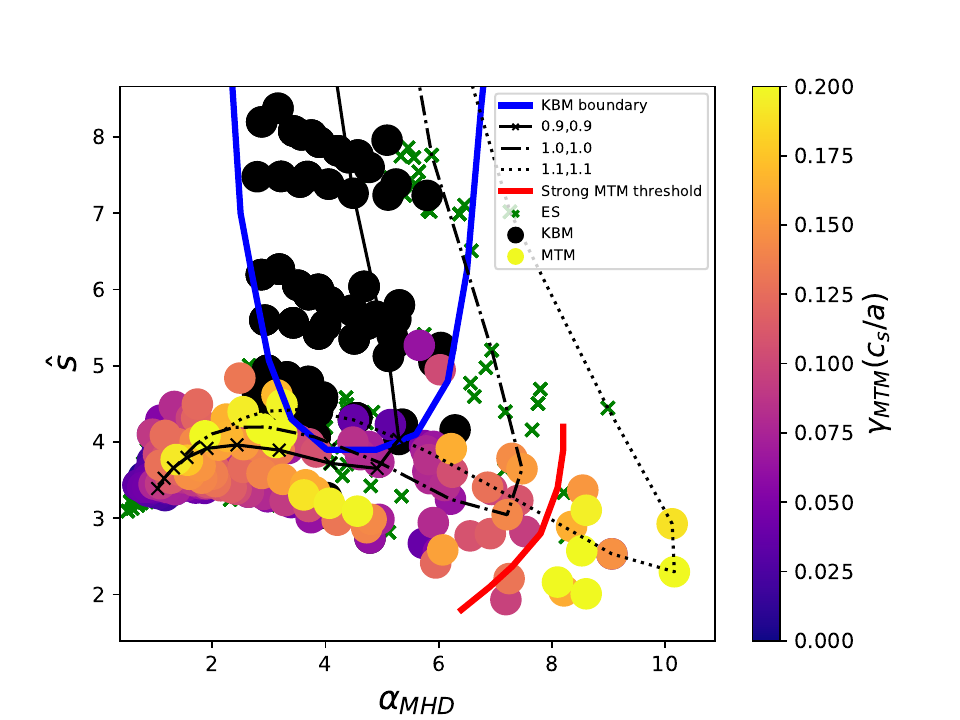}
    \caption{\label{s_alpha}  Instability and equilibrium information in $\hat{s}-\alpha$ space for local linear \textsc{GENE} simulations spanning the equilibria in all three discharges.  Strong KBM instabilities are shown in black while MTM is shown in color (denoting growth rate).  The blue curve roughly traces a ballooning stability boundary, while the red curve roughly traces a strong MTM threshold.  The black lines trace the radial points for three pressure gradient variations based on the 162940 case.  Note that the strong MTM threshold (red line) lies just beyond the pre-ELM equilibrium.    }
\end{figure}

In order to better understand the stability landscape of the pedestal, Fig.~\ref{s_alpha} shows information about the various instabilities on a familiar $\hat{s}-\alpha$ plot.  This plot utilizes information from the local linear gyrokinetic analysis of the ensembles of equilibria for all three discharges at radial locations spanning the pedestal.  Each point in the scatter plot corresponds to the $\hat{s}=\frac{\rho_{tor}}{q}\frac{d q}{d \rho_{tor}}$ and $\alpha$ at some radial position in one of the equilbria.  Here, $\alpha$ is defined as $\alpha = q_0^2 R \beta_e  \sum_s n_{0s} T_{0s} \left ( \omega_{Ts} + \omega_{ns} \right )$, where $s$ denotes the plasma species, the major radius $R$ is normalized to the reference scale length for the simulation (here the minor radius), and the local density and temperature are normalized to the electron density and temperature.

The MTMs are shown with colored symbols (denoting growth rates) identified by the criteria described above in Eqs.~\ref{QEM_criterion},\ref{parity_criterion},\ref{omega_criterion}.  The remaining instabilities are classified as strong KBMs ($\hat{E}_{||} \leq 0.2$, which would approximate ideal ballooning modes) and modes that are more electrostatic in nature ($\hat{E}_{||} > 0.2$, denoted with green $x$ symbols).

The strong KBMs cluster in the expected instability region at higher values of magnetic shear and moderate values of $\alpha$ (the blue line was added to roughly outline this ballooning-unstable region).  The MTMs at low $\alpha$ are unstable at the pedestal top (roughly $\rho_{tor}<0.95$).  The other cluster of MTMs (high $\alpha$, low $\hat{s}$) lies in a region of ballooning second stability, which corresponds to the radial positions with steep-gradients in the middle of the pedestal.  The three black lines (with various line styles) trace the magnetic shear and $\alpha$ for three different equilibria (for the 162940 discharge).  This shows that the pre-ELM fit (labeled 1.0 1.0) includes several points in ballooning second stability.  For equilibria with steeper gradients, the points extend even further into the second stability region, as indicated by the dotted line, which corresponds to the case with increased gradients in both temperature and density (labeled 1.1 1.1).  The red line was added to highlight a transition to strongly unstable MTMs (note the transition to the bright yellow symbols denoting high growth rates), which occurs just above the nominal pre-ELM equilibrium (i.e., between the 1.0 1.0 line and the 1.1 1.1 line).  {\it This suggests that the pre-ELM pedestal lies at a strong MTM threshold and MTM is likely responsible for constraining the pedestal by producing a threshold in the second stability window. }  In the next section, we further investigate whether MTM could play the role of an inter-ELM limit on the pedestal pressure (not merely temperature).

Before proceeding, we briefly discuss the concept in the literature of a so-called `ballooning-critical pedestal' (BCP) or `gyrokinetic-critical pedestal' (GCP)~\cite{Parisi_2024a,Parisi_2024b}.  According to this prescription, the pedestal is considered to be at the KBM threshold when KBM is unstable at half of the radial positions in the pedestal.  We were unable to identify such dynamics in the present analysis, in large part due to the cases with strongly unstable KBM at low gradients, which then transition to KBM stability at steeper gradients.  However, Fig.~\ref{s_alpha} does show some connections with such a picture: the dashed line denoting the $\hat{s}-\alpha$ points for the pre-ELM state (labeled 1.0 1.0) intersects the ballooning unstable region at the foot of the pedestal (high $\hat{s}$) and closely approaches the ballooning unstable region near the pedestal top (lower $\hat{s}$ and lower $\alpha$).  However, even in this scenario, the MTM appears necessary to close the second stability gap in the mid-pedestal.

\subsection{MTM particle flux and dependence on density gradients}
\label{sec:MTM_gradP}

Since the MTM is driven {\it primarily} by electron temperature gradients and produces {\it primarily} electron thermal transport, it has typically been viewed as a constraint on pedestal temperature but not pressure~\cite{kotschenreuther_19,TPT}.  We investigate and subtly revise these concepts in this section.   

\begin{figure}[H]
    \centering
    \includegraphics[scale=0.8]{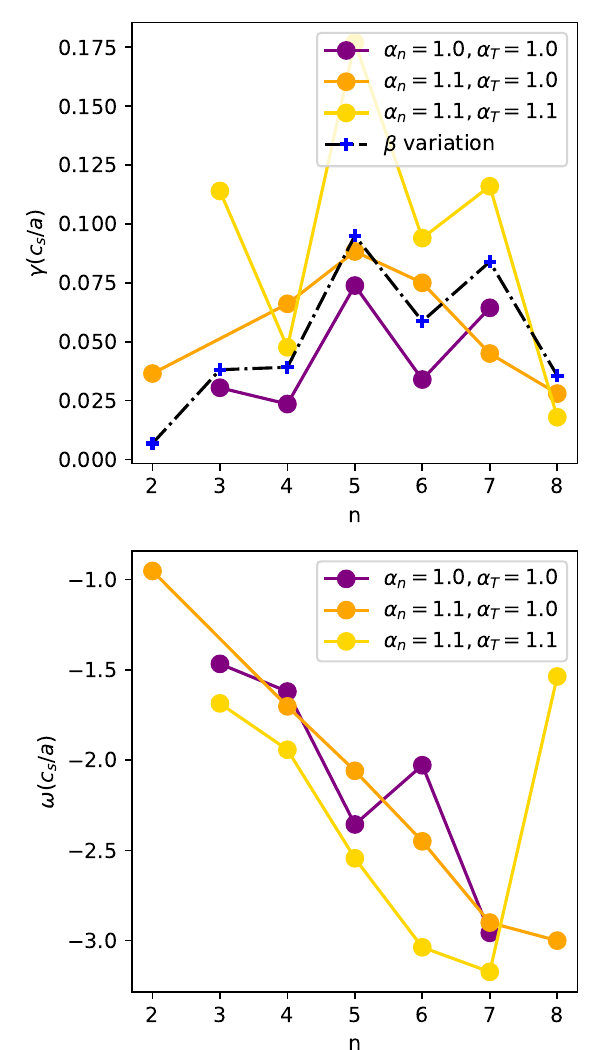}
    \caption{\label{gamma_omn} Growth rates and frequencies from global \textsc{GENE} simulations for three variations of temperature and density profiles.  Note that the MTM growth rates increase with density gradients, not just temperature gradients, indicating some degree of pressure gradient drive.  The blue symbols denote a simulation increasing $\beta$ from the base case to match the $\alpha_n = 1.1$ case.    }
\end{figure}

First, we examine the density gradient dependence of MTM growth rates by considering the global instabilities for three of the equilibria: the nominal pre-ELM fit ($\alpha_T = \alpha_n = 1.0$), the case with a $10\%$ increase in the density gradient ($\alpha_T = 1.0$ and $\alpha_n = 1.1$), and the case with both gradients increased ($\alpha_T = \alpha_n = 1.1$).  Fig.~\ref{gamma_omn} shows the growth rates and frequencies for these cases (note that this figure reproduces some of the data from Fig.~\ref{global162940}).  Clearly, the growth rates increase with density gradient alone.  {\it This suggests that in the pedestal parameter regime, MTMs can be driven by pressure gradients, not just temperature gradients}.  Note that, since these simulations are based on fully self-consistent equilibria, multiple parameter changes take effect with the increased density gradient including higher collisionality, lower magnetic shear, and higher Shafranov shift.  The interplay of these effects will be investigated in more depth in Sec.~\ref{sec:nsep}.   

\begin{figure}[H]
    \centering
    \includegraphics[scale=0.8]{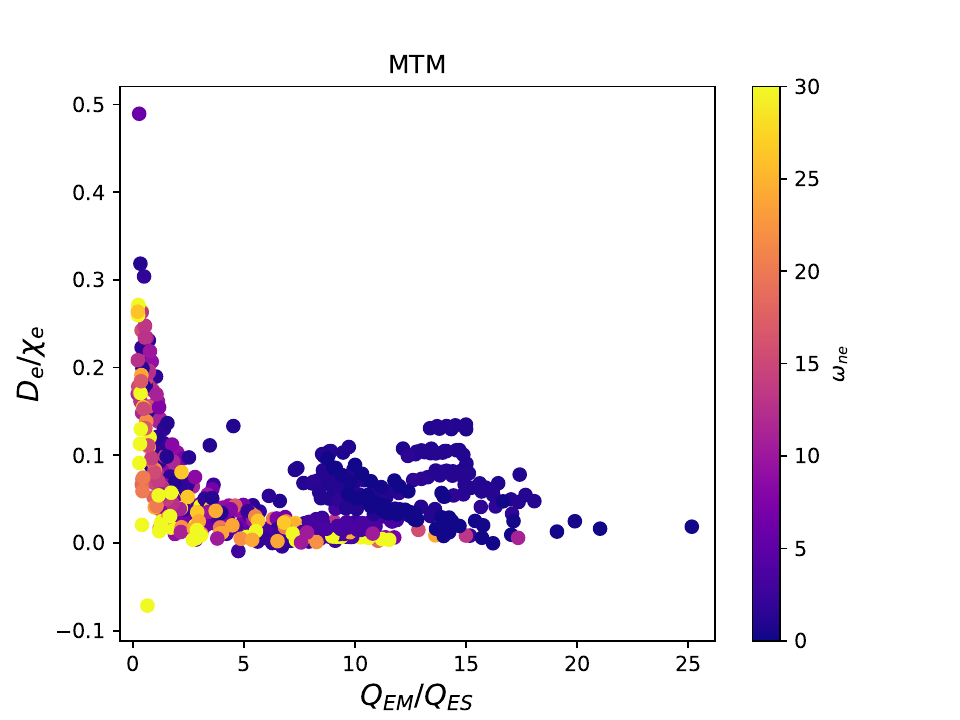}
    \caption{\label{D_chi}  The particle to thermal diffusivity ratio $D_e/\chi_e$ for a large number of MTMs spanning the three discharges.  The MTMs are identified according to the criteria defined in Eqs.~\ref{QEM_criterion},\ref{omega_criterion},\ref{parity_criterion}. Relatively strong particle transport is observed for many cases with steep density gradients and lower values of (normalized) electromagnetic heat flux.  Note that KBM would have much larger ratios of $D_e/\chi_e$~\cite{kotschenreuther_19}.      }
\end{figure}

As a second test, we survey the ratio of particle to thermal diffusivity ($D_e/\chi_e$) for the entire data set of local linear MTM.  Fig.~\ref{D_chi} shows this ratio plotted against $\hat{Q}_{EM}$.  To enrich the information of this plot, the density gradient $\omega_{ne}$ is shown in color.  At low values of $\hat{Q}_{eEM}$, the ratio of diffusivities can reach values up to 0.3, which is in the relevant experimental range identified by edge modeling~\cite{callen_10,TPT}.  For shot 162940, SOLPS modeling predicts values of $D_e/\chi_e\approx 0.05-0.25$ depending on radial position in the pedestal as well as various modeling assumptions~\cite{TPT}.  Note that all instabilities in this figure are selected using the MTM criteria defined above in Eqs.~\ref{QEM_criterion},\ref{parity_criterion},\ref{omega_criterion}.  The fact that the larger particle fluxes occur primarily at lower values of $\hat{Q}_{EM}$, suggests that perhaps the modes take on a hybrid nature encompassing some MHD-like properties.  Most of these high $D_e/\chi_e$ instabilities are produced by relatively steep density gradients (denoted by the color), indicating that these modes are produced in the steep gradient (second stability) region of the pedestal. 

To summarize, MTMs in the pedestal can be driven by {\it pressure} gradients and produce experimentally relevant particle transport, establishing them as a viable mechanism for constraining the pedestal {\it pressure} and not merely the pedestal temperature.   While these characteristics don't directly contradict predictions in Ref.~\cite{kotschenreuther_19}, they push the boundaries of expected MTM `fingerprints' and broaden the potential scope of the role of MTM in the pedestal.

\subsection{Mode Properties}
\label{sec:mode_properties}

This section describes several properties of the pedestal MTMs, notably, their dependence on ballooning angle, collisionality, and driving gradients.  

\begin{figure}[H]
    \centering
    \includegraphics[scale=0.8]{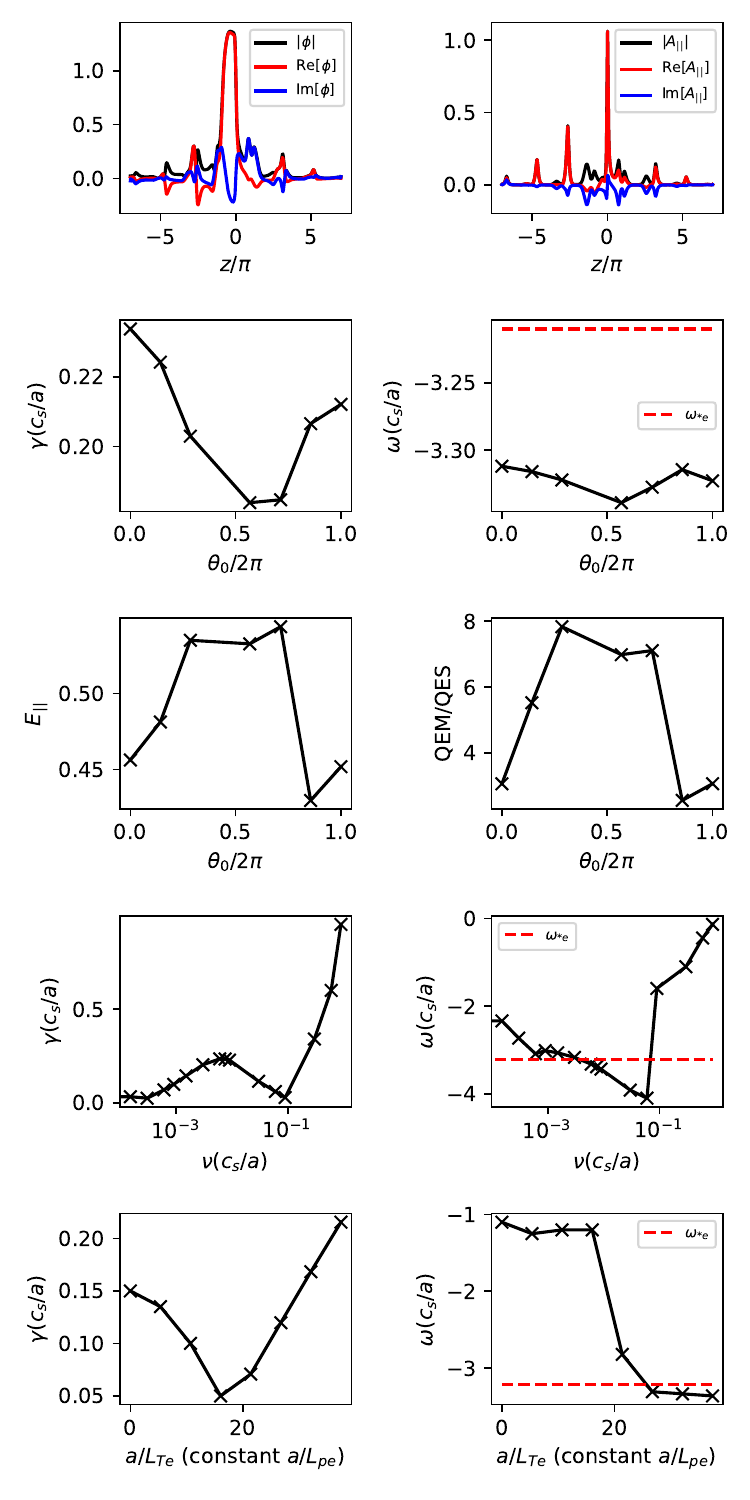}
    \caption{\label{mode_info1} Eigemodes (top panel), ballooning angle dependence (second and third panels), collisionality dependence (fourth panel) and temperature gradient dependence (at fixed pressure gradient---bottom panel) for a representative \textsc{GENE} simulation with $k_y \rho_s = 0.05$, $\rho_{tor} = 0.97$, and $\alpha_n = \alpha+T = 1.1$.   }
\end{figure}

\begin{figure}[H]
    \centering
    \includegraphics[scale=0.8]{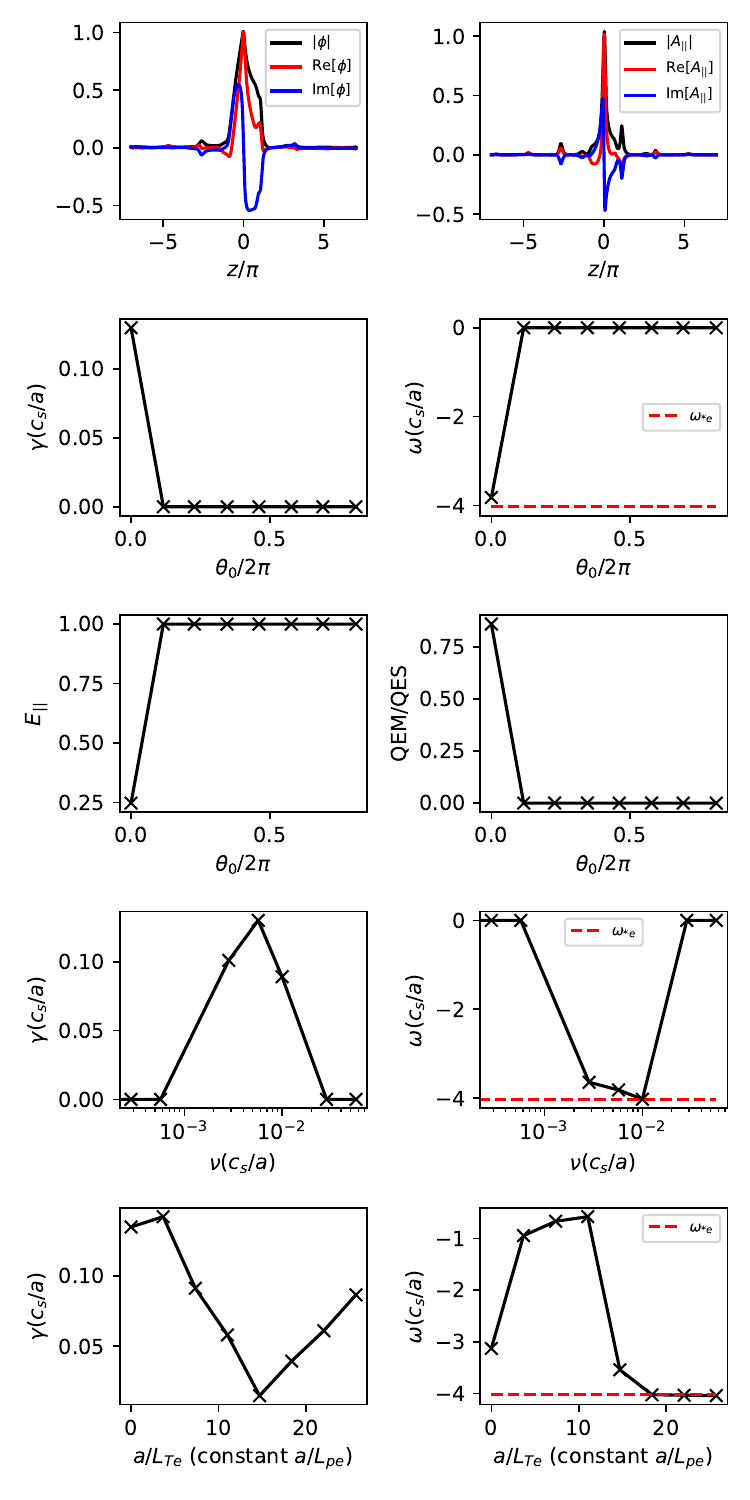}
    \caption{\label{mode_info2} Eigemodes (top panel), ballooning angle dependence (second and third panels), collisionality dependence (fourth panel) and temperature gradient dependence (at fixed pressure gradient---bottom panel) for a representative \textsc{GENE} simulation with $k_y \rho_s = 0.09$, $\rho_{tor} = 0.965$, and $\alpha_n = \alpha+T = 1.0$.   }
\end{figure}

\begin{figure}[H]
    \centering
    \includegraphics[scale=0.8]{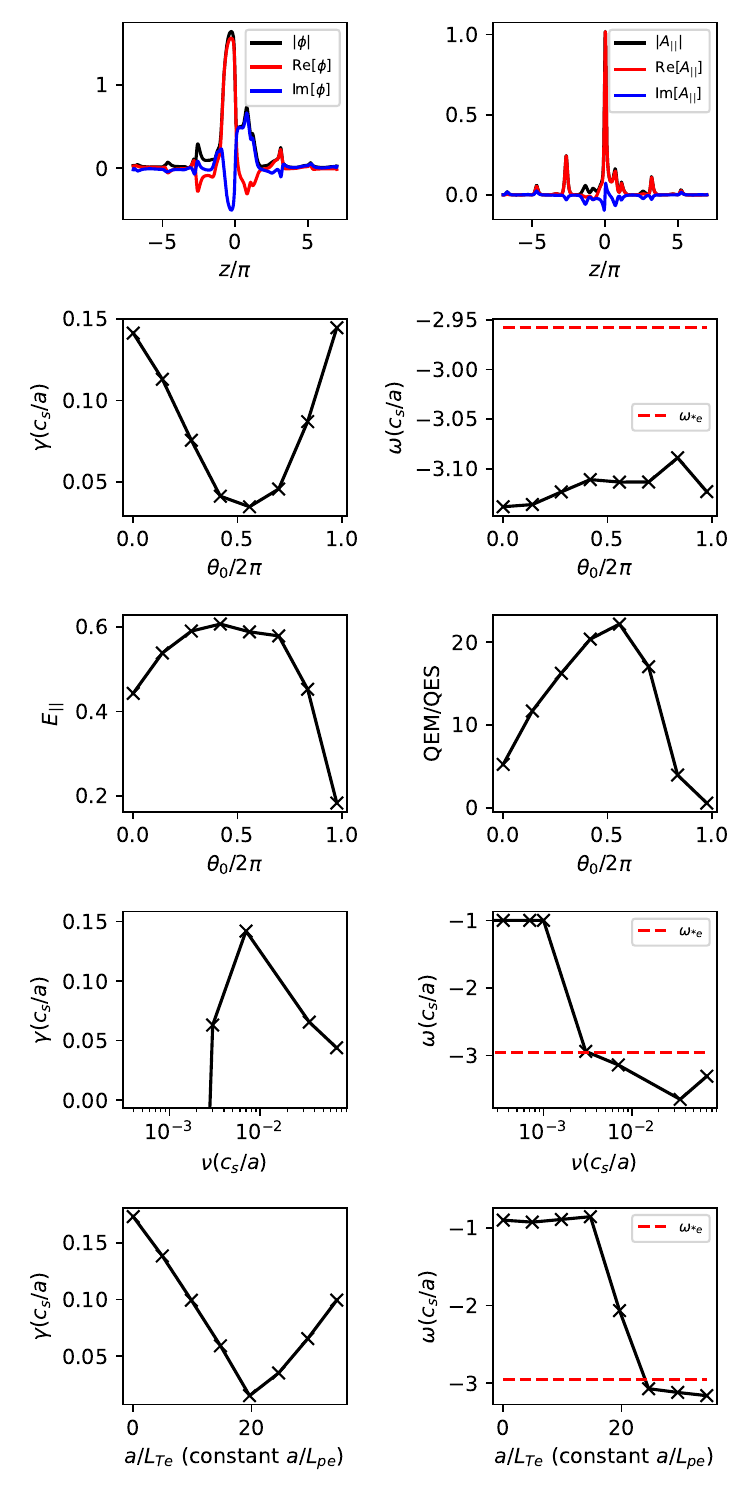}
    \caption{\label{mode_info3} Eigemodes (top panel), ballooning angle dependence (second and third panels), collisionality dependence (fourth panel) and temperature gradient dependence (at fixed pressure gradient---bottom panel) for a representative \textsc{GENE} simulation with $k_y \rho_s = 0.05$, $\rho_{tor} = 0.97$, and $\alpha_n = \alpha+T = 1.0$.   }
\end{figure}

Eigenmode information is plotted for three representative local linear simulations: $k_y \rho_s = 0.05$, $\rho_{tor} = 0.97$, $\alpha_n = \alpha_T = 1.1$ (Fig.~\ref{mode_info1}); $k_y \rho_s = 0.09$, $\rho_{tor} = 0.965$, $\alpha_n = \alpha_T = 1.0$ (Fig.~\ref{mode_info2}); and $k_y \rho_s = 0.05$, $\rho_{tor} = 0.97$, $\alpha_n = \alpha_T = 1.0$ (Fig.~\ref{mode_info3}).

{\bf Eigenmode structure:}  The MTMs all have `mixed parity' in the sense that the eigenmode structures for the electrostatic potential (top left plots) are not purely anti-symmetric and the eigenmode structures of the magnetic vector potential are not purely symmetric (top right plots).  In all cases, however, the magnetic vector potential has sufficient tearing parity to allow for an eigenmode averaged net radial magnetic field, and consequently, magnetic field line tearing.  

{\bf Ballooning angle dependence}:  The three cases exhibit substantial variation in their dependence on ballooning angle.  Fig.~\ref{mode_info1} shows very weak dependence, suggesting that the instability is largely slab-like, as also noted for low-$n$ pedestal MTM in Refs.~\cite{hassan_NF_21},~\cite{halfmoon_pop_22}.  However, Fig.~\ref{mode_info2} exhibits strong stabilization with ballooning angle, suggesting strong toroidal effects.  It's also notable that this example exhibits the lowest value of $Q_{eEM}/Q_{eES}$, possibly suggesting some degree of MHD-like behavior and a larger role for the electrostatic potential and particle transport (recall Fig.~\ref{D_chi}).  

{\bf Collisionality dependence}  All three examples exhibit the expected collisionality dependence of collisional MTM: a peak in growth rates at intermediate collisionality and full suppression in the collisionless limit (contrast with the collisionless branches of MTM discussed in Refs.~\cite{predebon_13,hassan_NF_21,halfmoon_pop_22} for different parameter regimes and/or higher wavenumber ranges).  

{\bf Gradient drive}  As noted in Sec.~\ref{sec:MTM_gradP}, the instability is---at least partially---driven by density gradients in addition to the conventional electron temperature gradient drive.  This is investigated in more depth with gradient scans shown in the bottom two plots in Figs.~\ref{mode_info1}-\ref{mode_info3}, which show scans of $\omega_{Te}$ keeping the pressure gradient fixed (i.e., $\omega_{Te} + \omega_{ne} = constant$).  In all cases, the growth rates go down substantially with $\omega_{Te}$, demonstrating that, while there may be some pressure gradient drive (as demonstrated in Fig.~\ref{gamma_omn}), the predominant thermodynamic drive is the electron temperature gradient, as must be true for MTMs.  In many cases, the MTM instability is replaced by something else at low values of $\omega_{Te}$, as determined by discontinuities in the mode frequency.  

In summary---the MTMs exhibit the salient features of the classical MTM instability: (1) frequencies near $\omega_{*e}$, (2) predominant electron temperature gradient drive, (3) sensitive dependence on collision frequency, (4) a substantial component of tearing parity in the magnetic vector potential, and (5) non-negligibale electromagnetic heat flux $\hat{Q}_{EM}$.  These features are consistent with the `fingerprint' picture described in Ref.~\cite{kotschenreuther_19}.  However, as outlined in this section as well as Sec.~\ref{sec:MTM_gradP}, other characteristics of the mode are somewhat unconventional, including: (1) mixed eigenmode parity, (2) a substantial contribution of electrostatic heat flux (sometimes), (3) a substantial contribution to particle flux (sometimes), and (4) some degree of density gradient drive (and by extension pressure gradient drive).  These novel characteristics allow for a new interpretation of the role of MTM in the pedestal.

\section{Quasilinear Mixing Length Diffusivities}
\label{sec:Dmix}

 Having investigated the linear physics in depth, we begin transitioning now to more concrete implications for transport and pedestal evolution.  A long-term goal of this work is to develop reduced models for pedestal transport that can be used for pedestal prediction.  To that end, we consider here quasilinear mixing length estimates for transport based on the linear simulations described above.  Profile predictions based on these mixing length estimates will be described in the next section.  The basic mixing length ansatz prescribes a diffusivity defined by the growth rate and perpendicular scale length of an eigenmode.  Here we use the following expression:

\begin{equation}
   \chi_{e,mix} = c_0 \max_{k_y} \frac{\gamma}{\langle k_\perp^2 \rangle}
\label{chimix}
\end{equation}
where the eigemode average ($<>$) is weighted by the electrostatic potential and $k_\perp$ incorporates both the radial and binormal wavenumbers as well as information from the magnetic geometry.  
 
 \subsection{Diffusivities for 162940}
 \label{sec:Dmix162940}

\begin{figure}[H]
    \centering
    \includegraphics[scale=0.8]{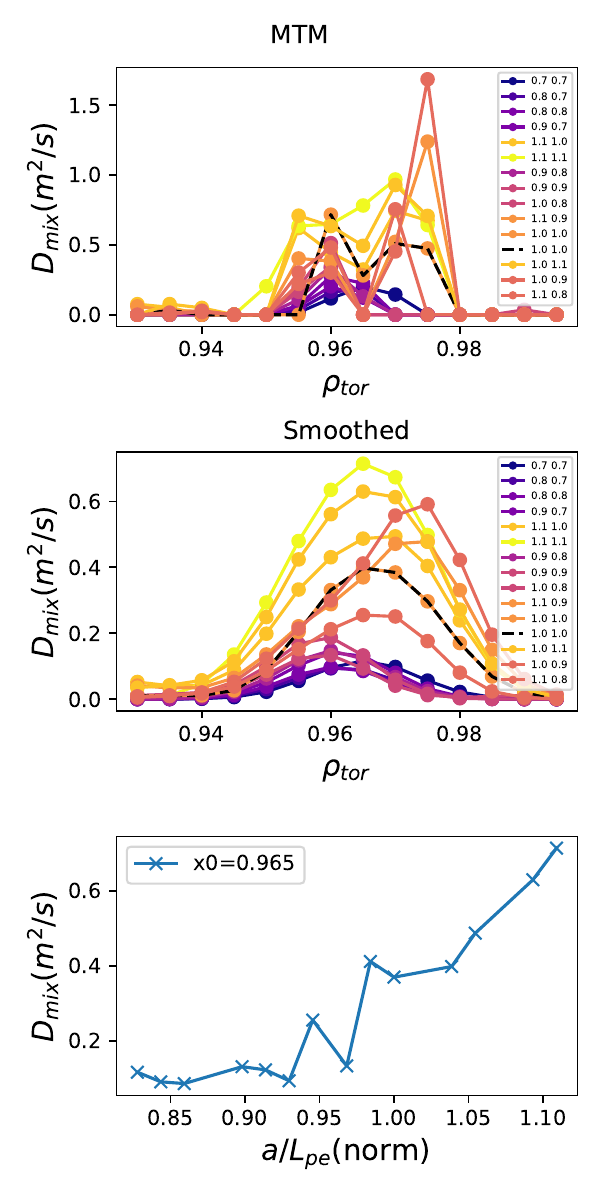}
    \caption{\label{Dmix_162940_MTM}   Mixing length estimates of thermal diffusivity spanning the pedestal for MTM instabilities for DIII-D shot 162940.  The middle panel shows a windowed smoothing of the top panel.  The bottom panel shows the diffusivity at $\rho_{tor}=0.965$ as a function of normalized pressure gradient. }
\end{figure}

\begin{figure}[H]
    \centering
    \includegraphics[scale=0.8]{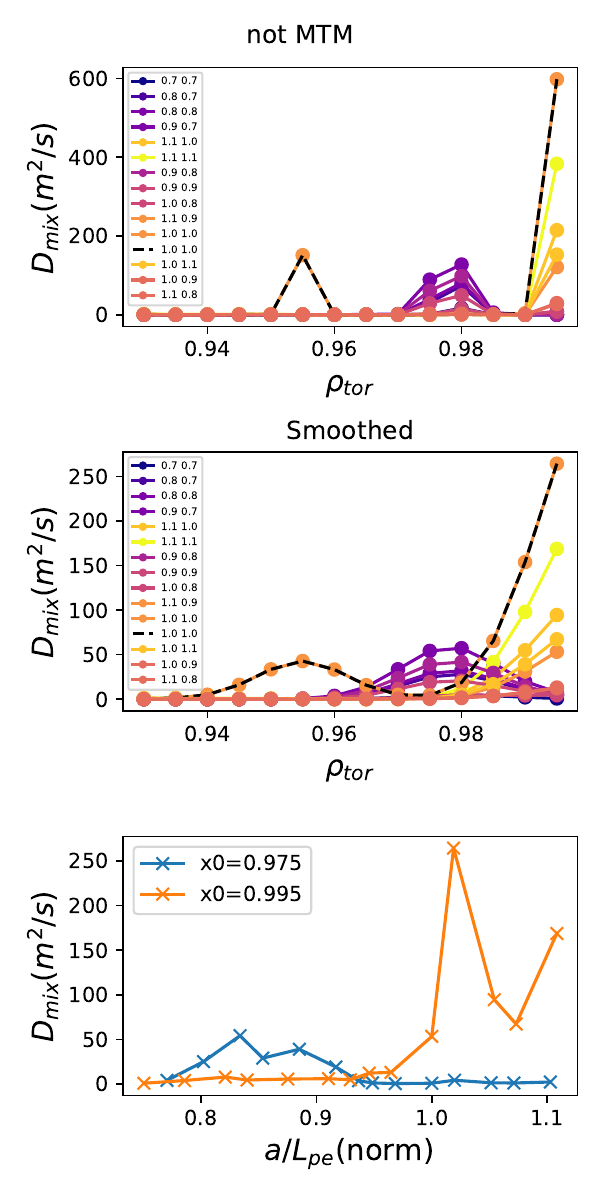}
    \caption{\label{Dmix_162940_notMTM}  Mixing length estimates of thermal diffusivity spanning the pedestal for all non-MTM instabilities for DIII-D shot 162940.  The middle panel shows a windowed smoothing of the top panel.  The bottom panel shows the diffusivity at $\rho_{tor}=0.975,0.995$ as a function of normalized pressure gradient. }
\end{figure}

\begin{figure}[H]
    \centering
    \includegraphics[scale=0.8]{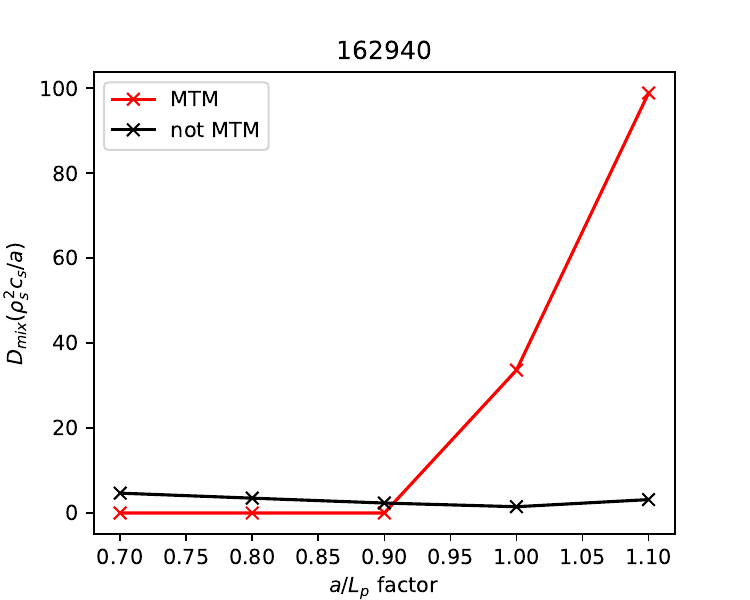}
    \caption{\label{global_Dmix_162940} Global mixing length diffusivities for DIII-D shot 162940 as a function of normalized pressure gradient for MTMs (red) and all other modes (black).   }
\end{figure}

Mixing length diffusivities for the shot 162940 are shown in Figs.~\ref{Dmix_162940_MTM} (for MTM), ~\ref{Dmix_162940_notMTM} (all modes except MTM), and ~\ref{global_Dmix_162940} (global).  For the local simulations, the top panel shows the quantity defined in Eq.~\ref{chimix} for radial positions spanning the pedestal.  Since there is substantial variation between neighboring points in the pedestal, we show also a windowed average of the mixing length estimates in the middle panel.  The bottom panel shows the progression of the diffusivity as a function of a `normalized' pressure gradient scale length.  Note that the global mixing length diffusivities are a rough estimate since the simulations are in real space and the $k_x$ averaging is not trivial.  As a simple proxy, the global estimate is simply $\gamma/k_y^2$.

The picture is consistent with the discussions above regarding MTM and KBM.  The MTM transport (Fig.~\ref{Dmix_162940_MTM}) is predominant in the mid-pedestal region ($\rho_{tor} \approx 0.95-0.975$), which is in second stability for ballooning modes for the steep gradient cases.  The KBM diffusivities (Fig.~\ref{Dmix_162940_notMTM}) clearly exhibit the transition to second stability: in the mid-pedestal, the diffusivities are extremely large for the weak gradient cases (purple) but KBM transport becomes negligible for steep gradients.  On the other hand, the outermost radial position shows the opposite behavior: transport increasing monotonically with pressure gradient.  This is consistent with Fig.~\ref{s_alpha}, which shows the outermost radial positions reentering the ballooning unstable region.  We hypothesize that this is likely a common feature of the pedestal, since $\alpha$ decreases and $\hat{s}$ increases sharply in close proximity to the separatrix (the latter due to the fact that $q \to \infty$ at the separatrix).  As expected from the linear global simulations, the MTM is the major instability in the mid-pedestal region and it increases monotonically with pressure gradient.  

\begin{figure}[H]
    \centering
    \includegraphics[scale=0.8]{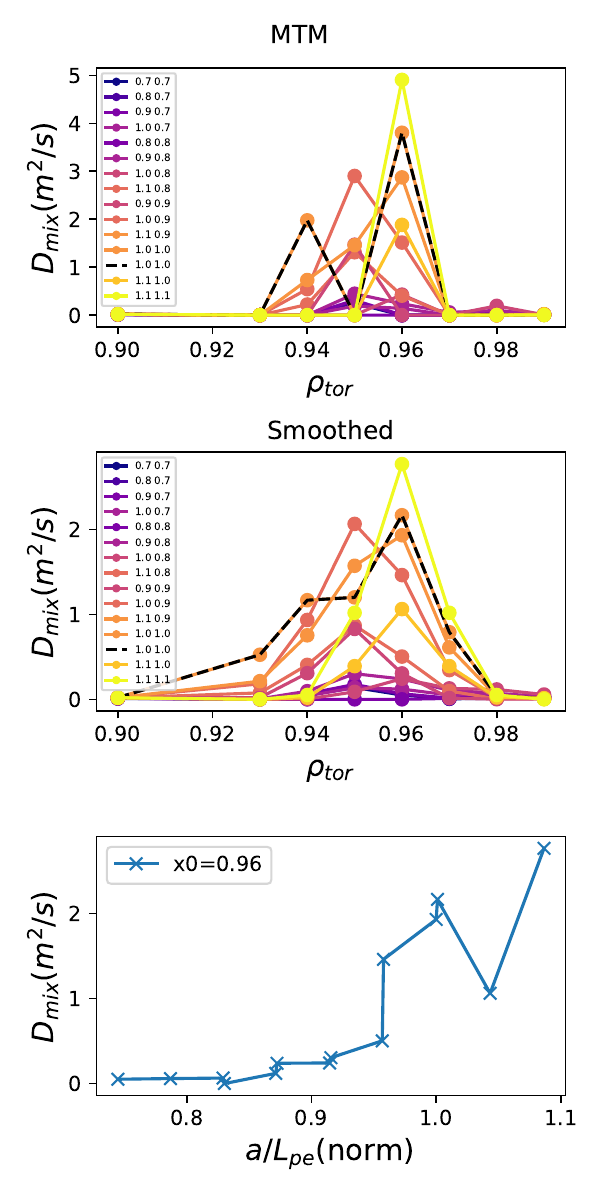}
    \caption{\label{Dmix_174082_MTM} Mixing length estimates of thermal diffusivity spanning the pedestal for MTM instabilities for DIII-D shot 174082.  The middle panel shows a windowed smoothing of the top panel.  The bottom panel shows the diffusivity at $\rho_{tor}=0.96$ as a function of normalized pressure gradient.   }
\end{figure}

\begin{figure}[H]
    \centering
    \includegraphics[scale=0.8]{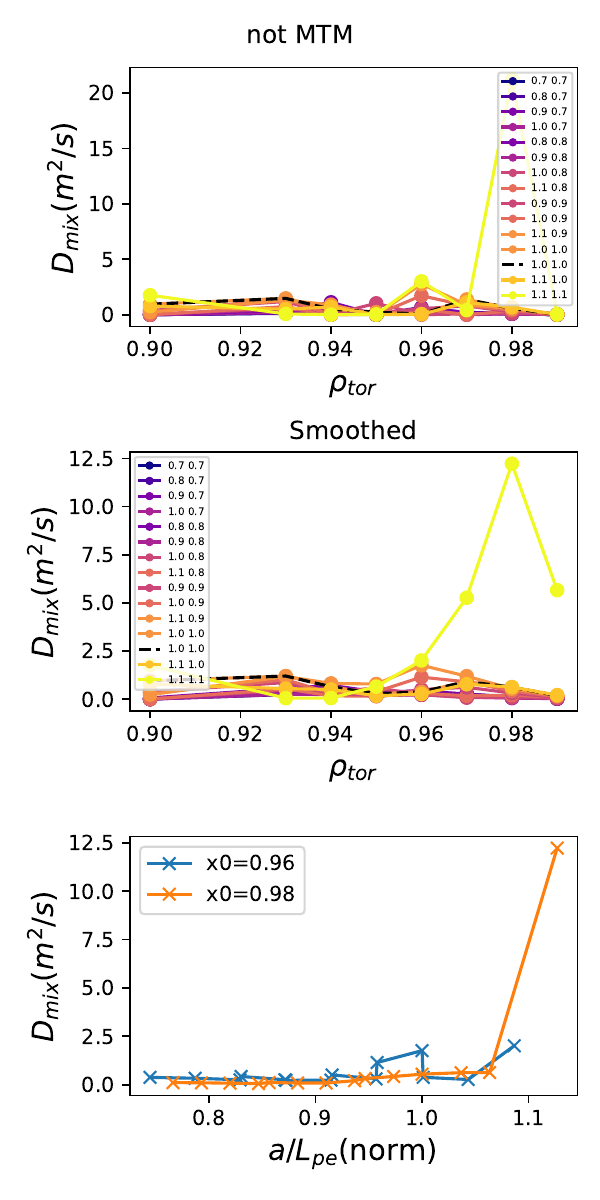}
    \caption{\label{Dmix_174082_notMTM} Mixing length estimates of thermal diffusivity spanning the pedestal for all non-MTM instabilities for DIII-D shot 174082.  The middle panel shows a windowed smoothing of the top panel.  The bottom panel shows the diffusivity at $\rho_{tor}=0.98$ as a function of normalized pressure gradient.    }
\end{figure}

\begin{figure}[H]
    \centering
    \includegraphics[scale=0.8]{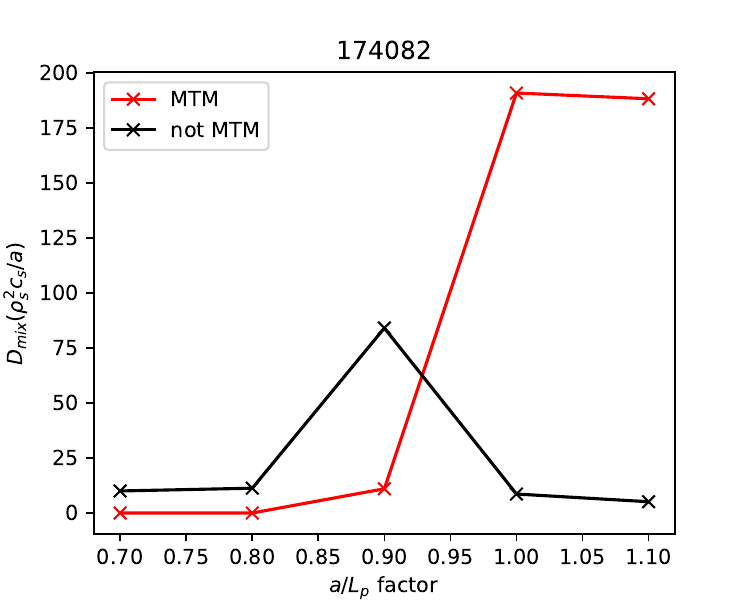}
    \caption{\label{global_Dmix_174082} Global mixing length diffusivities for DIII-D shot 174082 as a function of normalized pressure gradient for MTMs (red) and all other modes (black).  }
\end{figure}

The diffusivities for case 174082 exhibit similar behavior.  Note that the pedestal in this case is much broader, so one cannot directly compare radial positions between the discharges.  In this scenario, MTM produces increasingly large transport as the pressure gradients increase, particularly in the mid-pedestal region as seen in Fig.~\ref{Dmix_174082_MTM} at $\rho_{tor} \sim 0.96$.  KBM instabilities are rather weak with the exception of the outer pedestal where a sudden threshold is encountered for the steepest gradient case as seen in Fig.~\ref{Dmix_174082_notMTM} at $\rho_{tor} = 0.98$.  In the global simulations, MTM is clearly dominant and exhibits the expected threshold behavior, as seen in Fig.~\ref{global_Dmix_174082}.   

The growth rates and mixing length diffusivities from these two cases illustrate a plausible story of inter-ELM pedestal evolution.  As pressure gradients increase during an ELM cycle, the mid-pedestal region enters second stability and KBM is incapable of arresting pedestal evolution.  MTM fills this second stability gap and provides an inter-ELM pressure limit.  Due to the novel characteristics of this pedestal MTM instability (described in Sec.~\ref{sec:MTM_gradP}), it possesses the characteristics necessary to constrain the pedestal {\it pressure} not merely the temperature. 

KBM arises at the foot of the pedestal and may act as a near- or cross-separatrix pedestal constraint.  Both of these dynamics are manifest in the local simulations, but due to the separatrix cutoff enforced in the global simulations, only the MTM is manifest.  

\begin{figure}[H]
    \centering
    \includegraphics[scale=0.8]{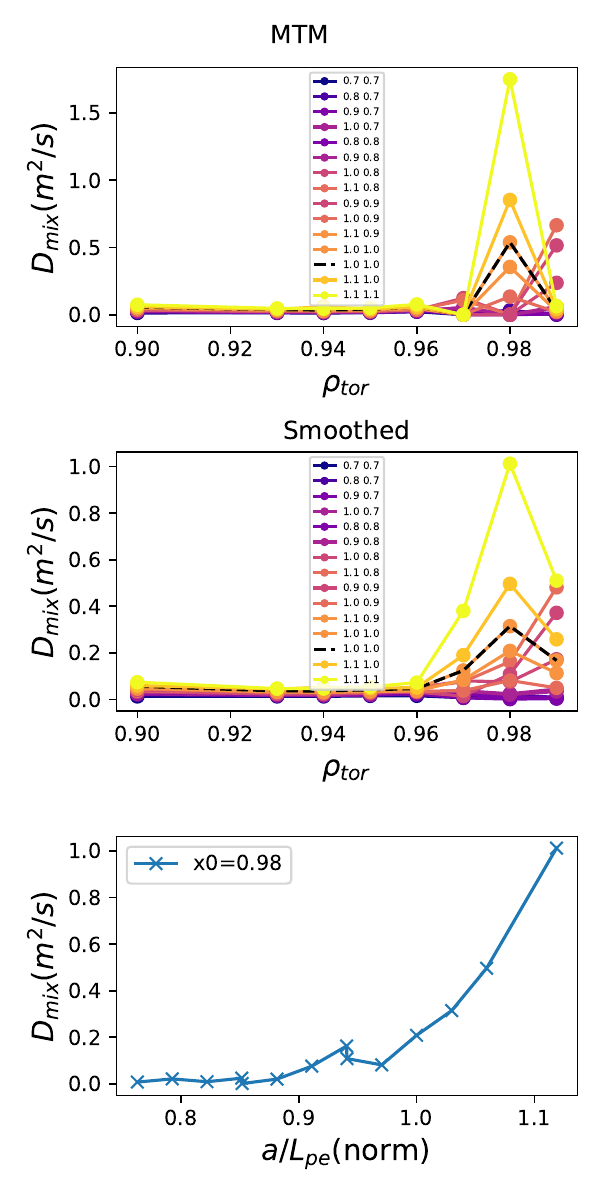}
    \caption{\label{Dmix_153764_MTM} Mixing length estimates of thermal diffusivity spanning the pedestal for MTM instabilities for DIII-D shot 153764.  The middle panel shows a windowed smoothing of the top panel.  The bottom panel shows the diffusivity at $\rho_{tor}=0.98$ as a function of normalized pressure gradient.   }
\end{figure}

The picture for 153764 exhibits some similarities and some contrasts with the other cases.  The main similarity is the activity of MTM in the mid-pedestal.  This is manifest clearly in the global simulations shown in Fig.~\ref{global153764}.  This shows predominant MTM transport at the steepest pressure gradient.  MTM activity is also clearly manifest in the local simulations, as shown in Fig.~\ref{Dmix_153764_MTM}, which shows clear threshold behavior in the mid-pedestal region.  However, the behavior of KBM is different from the other cases.  In 153764, KBM is unstable at the {\it same} radial position as the MTM and also exhibits threshold-like behavior.  Moreover, non-MTM instabilities (more electrostatic) are the predominant transport mechanism at the pedestal top.

In all cases, MTM appears to provide an inter-ELM pressure limit in the mid-pedestal region.  In two of the three cases, this appears to cut off a region of second stability for KBM, while KBM reaches a threshold at the foot of the pedestal.

\begin{figure}[H]
    \centering
    \includegraphics[scale=0.8]{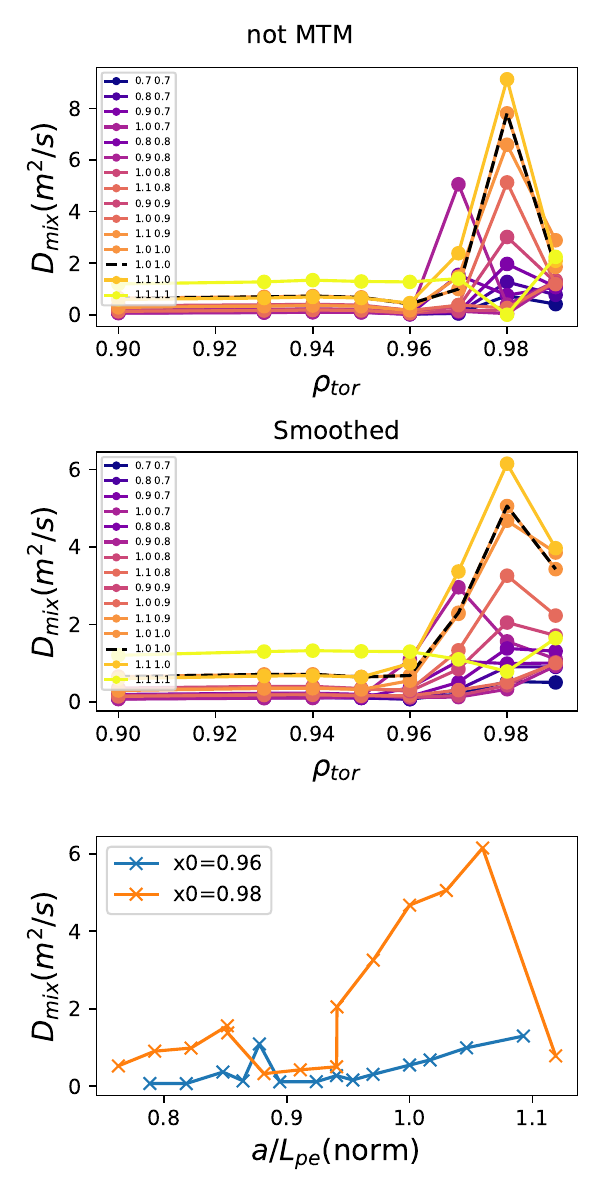}
    \caption{\label{Dmix_153764_notMTM} Mixing length estimates of thermal diffusivity spanning the pedestal for all non-MTM instabilities for DIII-D shot 153764.  The middle panel shows a windowed smoothing of the top panel.  The bottom panel shows the diffusivity at $\rho_{tor}=0.96,0.98$ as a function of normalized pressure gradient.  }
\end{figure}

\begin{figure}[H]
    \centering
    \includegraphics[scale=0.8]{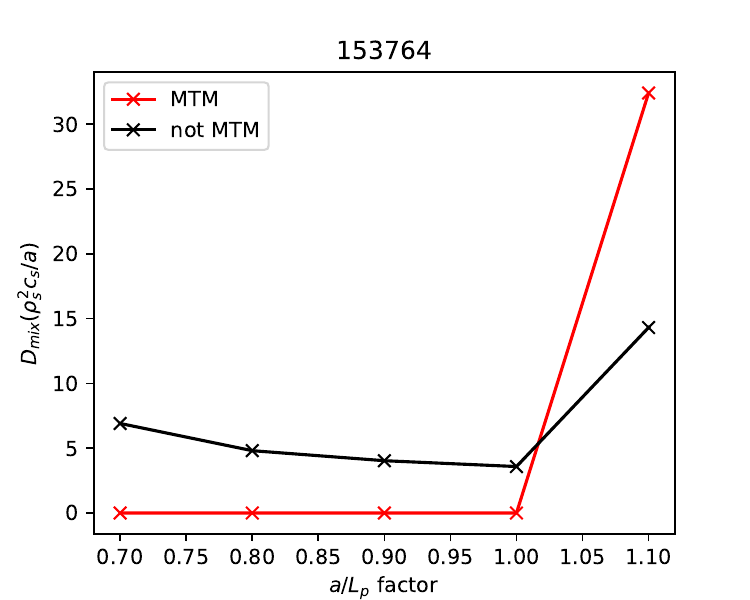}
    \caption{\label{global_Dmix_153764} Global mixing length diffusivities for DIII-D shot 153764 as a function of normalized pressure gradient for MTMs (red) and all other modes (black).   }
\end{figure}

\section{Transport Modeling for Pedestal Profile Evolution}
\label{sec:profile_predictions}

With the goal of predictive modeling, we use ASTRA to evolve pedestal profiles in response to the mixing length diffusivities described in the previous section.  Here for simplicity we focus on MTM for discharge 162940.  A more comprehensive treatment of all three discharges, encompassing both MTM and KBM, will be published elsewhere.  In order to enable efficient coupling with ASTRA, a discharge-specific surrogate model is constructed for the diffusivities by interpolating in the 3D parameter space of $\omega_{Te}$, $\omega_{ne}$, and $\rho_{tor}$.  Since the gyrokinetic simulations are based on full self-consistent equilibria for the fourteen sets of pedestal profiles, the other parameter dependencies (e.g., geometric information like safety factor and magnetic shear) are wrapped into this surrogate model automatically as long as the profile shape does not deviate too much from the underlying set of equilibria.  

\begin{figure}[H]
    \centering
    \includegraphics[scale=0.8]{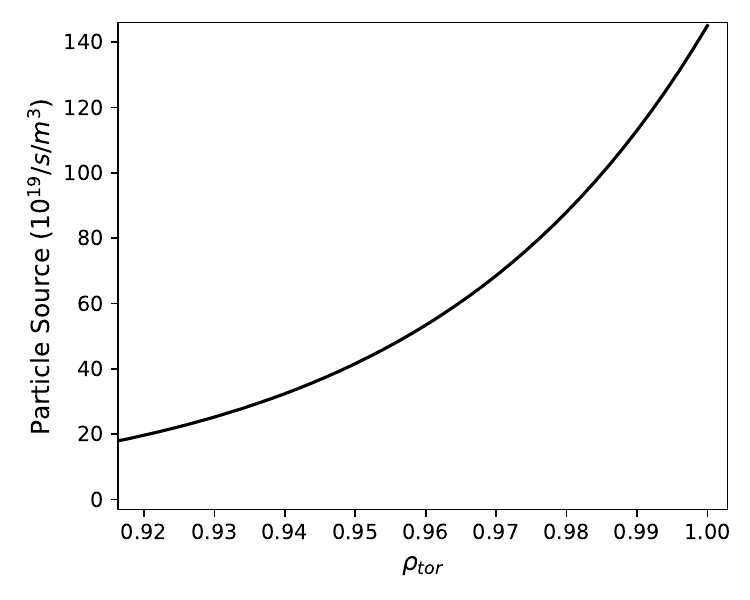}
    \caption{\label{particle source} Particle source estimate from SOLPS interpretive modeling for DIII-D shot 162940.   }
\end{figure}

The ASTRA simulations carried out for this study evolve both $T_e$ and $n_e$ profiles in response to neoclassical transport (supplied by NCLASS), ETG transport, and MTM transport supplied by the quasilinear mixing length model.  The ETG transport is supplied by a simple formula based on a large dataset of nonlinear ETG simulations in the pedestal~\cite{hatch_22,hatch_24}.  The particle source is estimated by interpretive edge modeling with SOLPS~\cite{guttenfelder_NF_21,TPT} and is shown in Fig.~\ref{particle source}.  The inter-ELM profiles reach a saturated state for a substantial fraction of the ELM cycle, so we run ASTRA to full saturation in order to compare with the experimental profiles.  

The model is defined in Eq.~\ref{chimix}.  The model has two free parameters, the prefactor, $c_0$, and the scale length for the Gaussian radial averaging.  The prefactor $c_0=3.0$ was selected to be in the range defined by local nonlinear simulations described in Ref.~\cite{hassan_NF_21}.  Due to the the sharp changes in the nature of the instabilities at closely neighboring radial positions, a radial smoothing function (Gaussian moving average) was applied to the diffusivities, which effectively adds one additional free parameter, namely the scale length of the Gaussian.  The effect of this averaging is shown in Figs.~\ref{Dmix_162940_MTM}~\ref{Dmix_153764_MTM}~\ref{Dmix_174082_MTM}.  This free parameter was adapted so that the ASTRA simulation matches the experimental data.  Due to this parameter tuning, this modeling exercise should be considered exploratory rather than predictive.  Ongoing work, which will be published elsewhere, is developing a more comprehensive transport model including $E \times B$ shear and testing this model across a broader set of scenarios (in addition to the entire set of DIII-D discharges described in this paper).  For particle transport, the particle diffusivity is defined by the quasilinear ratio of the relevant gyrokinetic eigenmode: $D_{e,mix} = \frac{D_{e,mode}}{\chi_{e,mode}}\chi_{e,mix}$.  

\begin{figure}[H]
    \centering
    \includegraphics[scale=0.8]{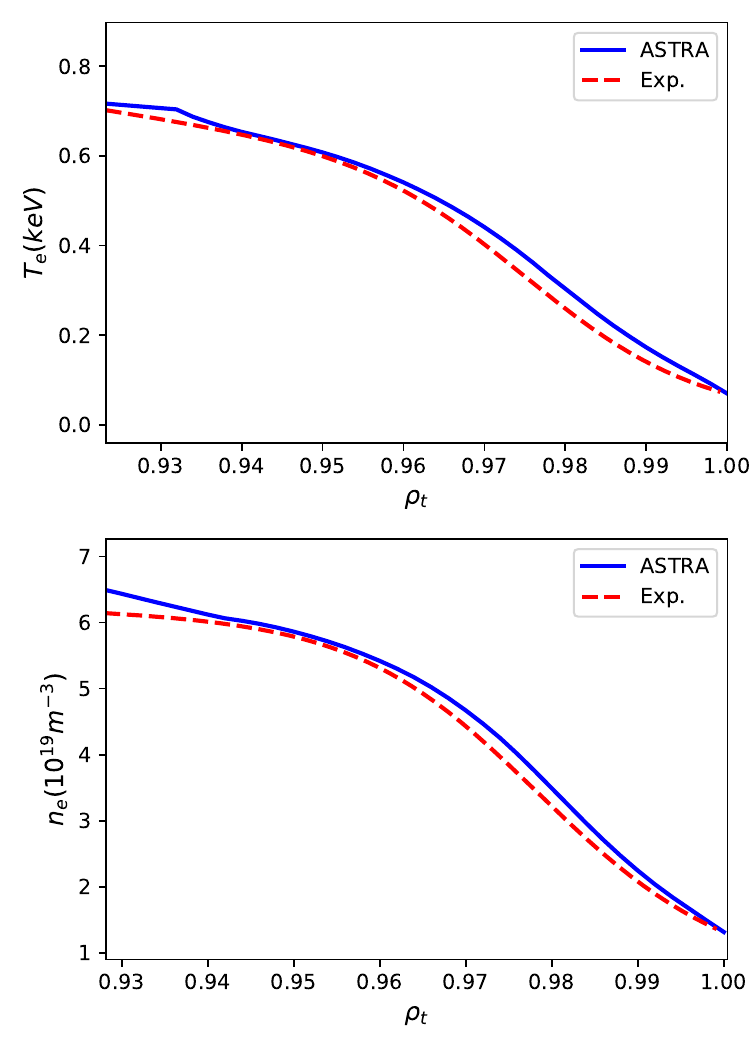}
    \caption{\label{profiles_astra1} Experimental pre-ELM profiles (red dashed) for DIII-D shot 162940 along with the saturated ASTRA simulations (blue) based on the quasilinear mixing length model along with models for neoclassical and ETG transport.   }
\end{figure}

Results of the profile predictions are shown in Fig.~\ref{profiles_astra1}, which exhibits extremely good agreement between modeling and experiment.  Note that the width of the radial smoothing factor was used as a free parameter to match the experiment.  However, the baseline transport level (i.e., the $c_0$ parameters) is consistent with nonlinear simulations.   Moreover, the fact that the modeling is able to simultaneously reproduce density and temperature is a strong test of self-consistency between the model and experiments.  The relative transport in the particle and thermal channels is entirely defined by the quasilinear ratio of the eigenmode---i.e., there are no free parameters whatsoever for this aspect of the modeling.  

\subsection{Sensitivity to SOL parameters}
\label{sec:nsep}

\begin{figure}[H]
    \centering
    \includegraphics[scale=0.8]{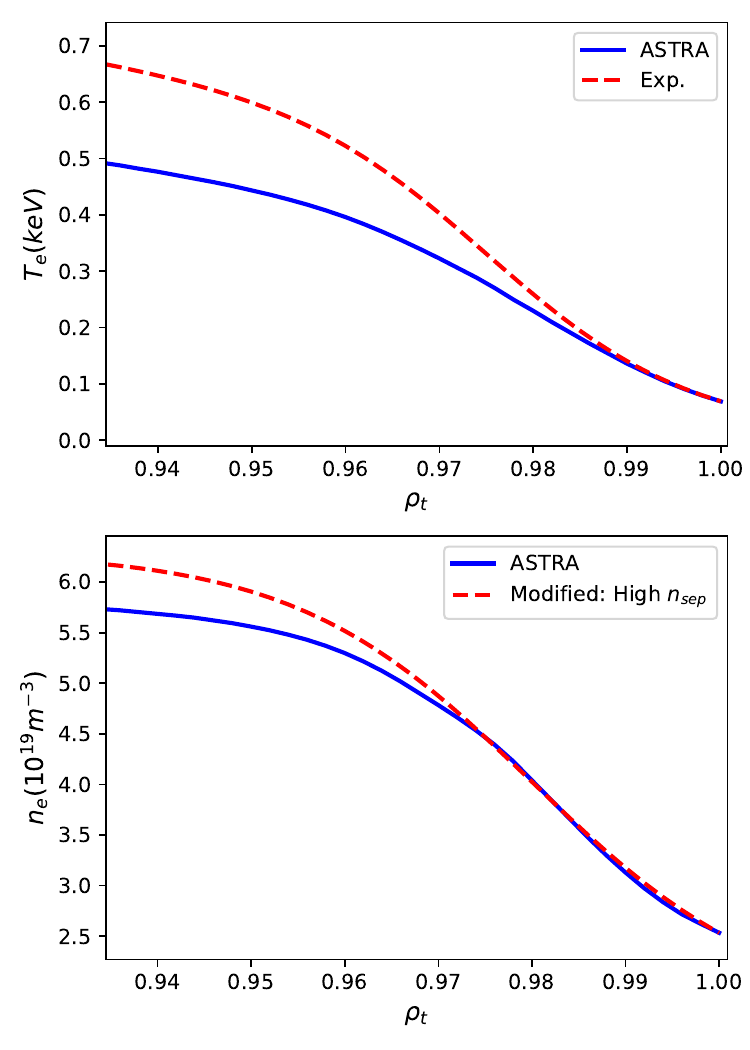}
    \caption{\label{profiles_astra2} Experimental pre-ELM profiles (red dashed) for DIII-D shot 162940 (the density has been modified for higher $n_{sep}$ while retaining the same pedestal top value) along with the saturated ASTRA simulations (blue) using the same model as shown in Fig.~\ref{profiles_astra1}.  The decreased pedestal pressure indicates reduced confinement with increased $n_{sep}$.   }
\end{figure}

One of the key effects for core edge integration is the strong impact of SOL parameters on core confinement.  Empirical evidence, as well as theory and simulation, points to a strong adverse effect from higher separatrix density, $n_{sep}$, on confinement~\cite{Maingi2010,maingi_15,stefanikova_18,frassinetti_21,lomanowski_22,hatch_17,hatch_19,kotsch_itpa}.   Currently, there is no model that can capture this sensitive connection.  As an initial application of this new modeling capability, we probe the interaction between $n_{sep}$ and the pedestal by applying the same ASTRA simulation setup to a scenario with higher $n_{sep}$ (an increase $n_{e,sep}=1.3 \rightarrow 2.5 (10^{19} m^{-3})$).  This requires an additional set of 14 equilibria with corresponding gyrokinetic simulations to parameterize a surrogate model for this scenario.  Results are shown in Fig.~\ref{profiles_astra2} keeping all parameters the same (aside from $n_{sep}$) as the baseline scenario.  This results in a substantial reduction of predicted $T_e$ as well as a modest reduction in pedestal top $n_e$.  The impact on confinement can be estimated from the relative pedestal top pressure, which exhibits a $32\%$ reduction for the high $n_{e,sep}$ case.  A recent analysis of the effect of separatrix density on confinement~\cite{kotsch_itpa} based on AUG data from the most recent ITPA H-mode confinement database~\cite{verdoolaege_21}, shows a roughly $50\%$ decrease in confinement over a similar range of $n_{sep}/n_{ped}$.  This is qualitatively consistent with the present modeling, assuming the reduction in pedestal top pressure is proportional to total stored energy.           

\begin{figure}[H]
    \centering
    \includegraphics[scale=0.8]{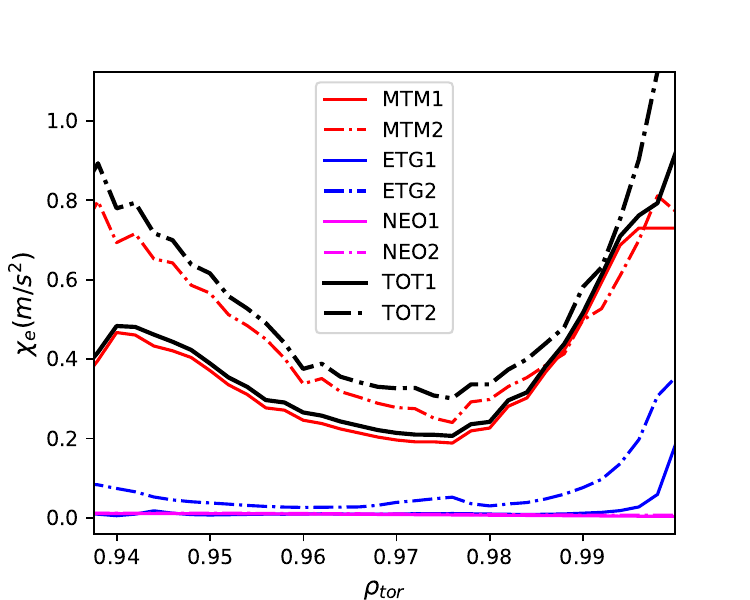}
    \caption{\label{chie_all} Contributions from MTM, ETG, and neoclassical to electron thermal diffusivity from ASTRA simulations for the baseline and high $n_{sep}$ cases.   }
\end{figure}

The diffusivities for this high $n_{sep}$ case along with those for the base case are shown in Fig.~\ref{chie_all}, revealing the following:  (1) a substantial increase in total thermal diffusivity, (2) a substantial increase in thermal diffusivity from MTM, and (3) a transition to a parameter regime where ETG becomes non-negligible, in contrast with the baseline scenario for which ETG is only active near the separatrix.  The ETG transport emerges due to a weaker density gradient.  The ETG transport scales like $\chi_e \propto \eta^{1.5}$ ($\eta = \omega_{Te}/\omega_{ne}$), making this transport mechanism very sensitive to variations in the density gradients.  This is a manifestation of the physics described in Ref.~\cite{kotschenreuther_24}, in which density gradient stabilization is investigated based on fundamental considerations of the gyrokinetic system.  This effect of separatrix density on the pedestal via ETG transport has been proposed as an explanation for degradation of pedestal performance for JET-ILW~\cite{hatch_17,hatch_19,chapman_21,chapman_25}.

\begin{figure}[H]
    \centering
    \includegraphics[scale=0.8]{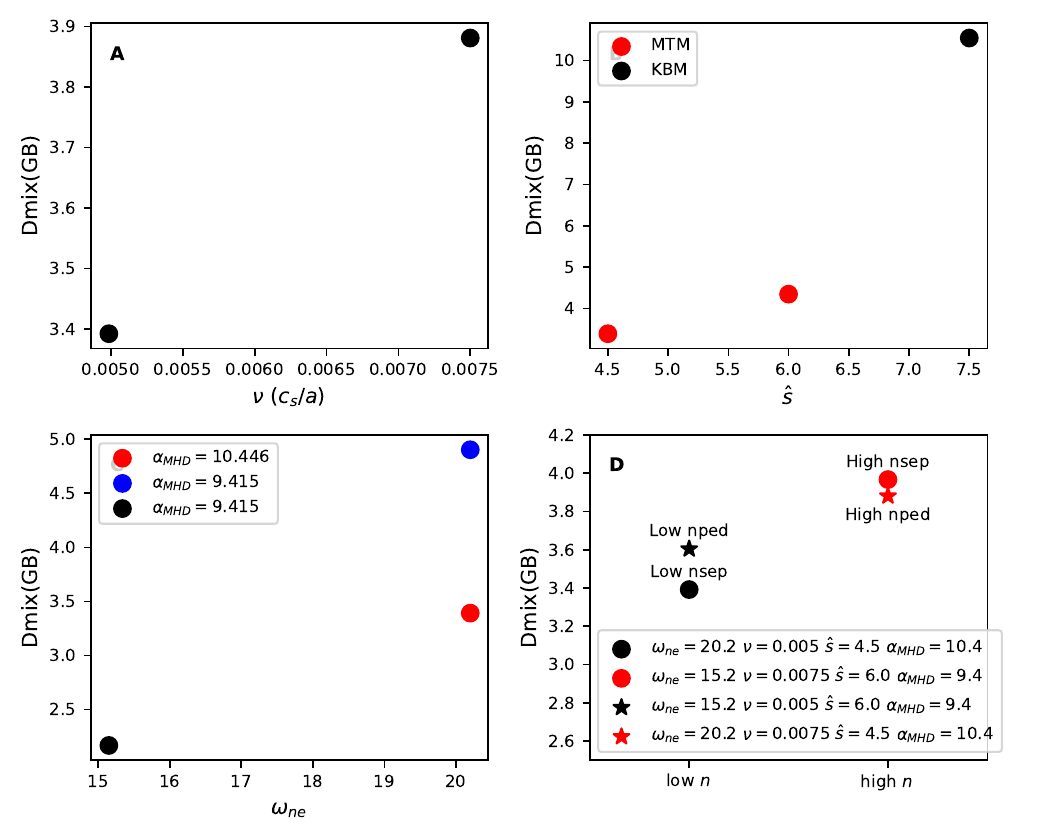}
    \caption{\label{nsep_nped_test} Mixing length diffusivities (maximized over $k_y$) from local linear \textsc{GENE} simulations using Miller geometry, exploring the isolated and combined effects of $\omega_n$, $\alpha_{MHD}$, $\hat{s}$, and collisionality. All instabilities are MTMs unless explicitly noted otherwise.  The mixing length diffusivity increases with collisionality, as seen in A.  Fig. B demonstrates that diffusivities increase with magnetic shear until KBM is destabilized.  Fig. C shows that diffusivities increase substantially with $\omega_n$; if $\alpha_{MHD}$ is adapted self consistently (red symbol), the diffusivity still increases albeit more weakly.  Fig. D shows the combination of these effects for two adaptations: (1) an increase in pedestal top density, and (2) an increase in separatrix density.  For the increase in pedestal top density, two of the four parameters are destabilizing.  For the increase in separatrix density, three of the four parameter changes are destabilizing.  In both cases, the net effect is an increase in transport.        }
\end{figure}

The mechanisms underlying the increase in MTM transport are more complex.  In order to investigate this in more depth, we carried out a series of local linear scans using Miller geometry, which allows for independent variation of all input parameters.  For density gradient variations, there are four main parameter dependencies: $\omega_n$, $\hat{s}$, $\alpha_{MHD}$, and collisionality $\nu$, which are explored in Fig.~\ref{nsep_nped_test}.  An increase in $n_{sep}$ (for fixed $n_{ped}$) would result in decreased density gradient, increased collisionality, increased magnetic shear, and decreased $\alpha_{MHD}$.  Fig.~\ref{nsep_nped_test} A shows an increase of the diffusivity with collisionality at this parameter point (although this can not be generalized universally due to the non-monotonic interplay between MTM stability and collisionality).  Fig.~\ref{nsep_nped_test} B shows an increase with magnetic shear, and Fig.~\ref{nsep_nped_test} C shows an increase with density gradients (partially mitigated by the corresponding increase in $\alpha_{MHD}$).  In other words, three of the four parameter changes produce increased diffusivity, resulting in a substantial net increase in transport for high $n_{sep}$ as shown in Fig.~\ref{nsep_nped_test} D.  For reference, we also investigate the corresponding result for an increase in $n_{ped}$ at fixed $n_{sep}$, which follows the same parameter trends with the exception of collisionality.  Although only two of the four parameter changes are destabilizing for the increase in $n_{ped}$, the net result is also a slight increase in diffusivity, which is also shown in Fig.~\ref{nsep_nped_test} D.  


\section{Discussion}
\label{sec:discussion}

One of the main claims of this paper is that MTM is responsible for setting an inter-ELM pressure limit in the mid-pedestal region where KBM is second stable.  In this section, we discuss how this fits into a broader pedestal picture as well as next steps for model development.  

The conventional picture of pedestal structure ascribes the role of inter-ELM pressure limit to KBM.  Our initial hypothesis at the outset of this work was consistent with this picture: KBM would be primarily responsible for particle transport and supply a pressure limit for the inter-ELM pedestal.  However, this picture is problematic for pedestals that lie in second stability where KBM is entirely stable.  This issue has been acknowledged in the literature.  One proposed resolution of this issue is to appeal to global KBM~\cite{saarelma_17}.  While we have not provided dispositive evidence against this theory, we find no evidence for it in our simulations.  In fact, KBM is more stable in the global simulations than local simulations.   However, the simulations described in this paper do not span the separatrix and hence do not include the vacuum region to the wall.  Consequently, we cannot rule out a potential role of global KBM.  We do propose, however, that the balance of evidence lies in favor of MTM as the inter-ELM pressure limit---at least for these discharges---based on the following observations: 
\begin{itemize}
    \item MTM exhibits a threshold at or near the pre-ELM state in all three discharges. 
    \item Experimentally observed frequencies in the magnetic diagnostics are consistent with MTM frequencies and not KBM frequencies.  In two cases, extremely close agreement was found between simulated frequencies and the fluctuations~\cite{hassan_NF_21,halfmoon_pop_22}.  In the third case, agreement was established within reasonable uncertainties~\cite{kotschenreuther_19}.  
    \item For discharge 153764, inter-ELM profile saturation correlates with the emergence of MTM frequencies~\cite{diallo_15,kotschenreuther_19}.
    \item Pedestal performance is correlated with magnetic fluctuations identified as MTM~\cite{chen_20}.  

\end{itemize}

The revised hypothesis from this work is that MTM provides the inter-ELM pressure constraint in the mid-pedestal region where KBM is stable.  KBM is likely active near the separatrix where magnetic shear is high.  KBM may also be active at the pedestal top, in which case a BCP (or GCP) picture may apply in combination with MTM closing off the second stability window.  

This work also points to open questions regarding the ELM trigger.  For the discharges studied, the inter-ELM profiles are saturated for a substantial fraction of the ELM cycle, so the ASTRA modeling described in Sec.~\ref{sec:profile_predictions} is a meaningful exercise.  However, this does not address the question of an ELM trigger.  Is the MTM a sort of kinetic precursor to the peeling ballooning instability (perhaps analogous to KBM vs. ideal ballooning)?  Or do MTMs trigger an ELM crash when a resonance condition is satisfied as proposed in Refs.~\cite{diallo2018direct,dominski2020identification}?  These questions will be investigate further in future work.

Parallel work has investigated this same quasilinear framework applied to NSTX~\cite{li_26}.  In that work, $T_e$ and $T_i$ profiles were reproduced accurately for two NSTX pedestals by coupling neoclassical transport, ETG transport, and KBM transport (from quasilinear gyrokinetics) with ASTRA using only one free parameter.  

Clear next steps include more extensive and comprehensive development and application of these reduced models.  Ongoing and future work includes testing and adding physics capabilities to the model, including incorporation of all instability types, addition of $E \times B$ shear, and coupling with peeling ballooning stability.  Such models are being deployed for development and validation across a much broader experimental dataset within a Bayesian framework for uncertainty quantification.  

One aspect of MTM physics that has been neglected in this work is microtearing modes at higher toroidal mode number, which are unstable at finite ballooning angle~\cite{hassan_NF_21,halfmoon_pop_22}.   


\section{Summary and Conclusions}
\label{sec:summary}

This paper uses global and local linear gyrokinetic simulations of ensembles of pedestal equilibria to provide new insight into pedestal evolution and its impact on confinement.  A total of 42 equilibria were investigated in detail---fourteen equilibria for each of three DIII-D discharges.  MTMs and KBMs were the most common instabilities found in the simulations.  A clustering algorithm was used to verify and guide the following physics-based metrics for distinguishing MTMs: $Q_{eEM}/|Q_{eES}+Q_{iES}|>0.2$, $\omega/\omega_{*e} < -0.5$, and parity $P(A_{||})>0.15$.  

The main conclusions of the analysis follow:
\begin{itemize}
\item MTM is near a stability boundary in the middle of the pedestal for all three scenarios and it exhibits threshold behavior---i.e., the growth rates (and transport estimates) increase with pressure gradients at and beyond the pre-ELM experimental profiles. 

\item  KBM is often second-stable in the mid pedestal due to the low magnetic shear and steep pressure gradients ($\alpha$) in this region.  Consquently, it is incapable of providing a pressure limit over much of the pedestal.  KBM does, however, often exhibit threshold behavior at the foot of the pedestal where magnetic shear remains high.  

\item The above points suggest that MTM and KBM may work in tandem to define the ultimate pressure limit spanning the entire pedestal.

\item The MTMs that are characteristic of the steep-gradient region of the pedestal have all the signatures of the conventional MTM instability plus some novel features:  higher particle transport than is typical for MTM, and some degree of density gradient drive (and by extension pressure gradient drive).  The MTM thus possesses the properties necessary to limit the pedestal {\it pressure}, not just the pedestal $T_e$.

\item Quasilinear mixing length estimates of MTM transport were used to evolve pedestal profiles (coupled to ASTRA).  With properly tuned free parameters, the model was capable of closely reproducing pedestal profiles of both temperature and density.

\item When the transport model was applied to a scenario with a $2\times$ increase in separatrix density (otherwise identical), the pedestal pressure decreased at a level consistent with empirical observations documented in the ITPA H-mode confinement database.  This confinement degradation can be attributed to increases in both MTM and ETG transport.  The increase in ETG transport was due to a decreased density gradient.  The increased MTM transport is attributable to the interplay of multiple effects, increased collisionality, decreased Shafranov shift, and increased magnetic shear.    

\end{itemize}

This paper sheds new light on the nature of pedestal instabilities and their role in constraining the pedestal.  It illustrates physical mechanisms that connect pedestal properties (and net confinement) with separatrix properties, which is perhaps the crucial question for core-edge integration moving toward a new generation of burning plasmas experiments, and ultimately fusion pilots plants.  It also lays the foundation for nascent predictive modeling capabilities using gyrokinetic-based reduced models.  

{\em Acknowledgements.--} 

This research used resources of the National Energy Research Scientific Computing Center, a DOE Office of Science User Facility.  We would like to thank Emiliano Fable, Giovanni Tardini, and Clemente Angioni for support with the ASTRA code.  This work was supported by U.S. Department of Energy Contract numbers: DE-SC0022164, DE-SC0024425, DE-FG02-04ER54742, DE-SC0022115.  D. R. Hatch and S. M. Mahajan report a financial and
leadership relationship with ExoFusion, a fusion-energy
company. This relationship did not influence the design,
execution, analysis, or interpretation of this research.
The terms of this relationship have been reviewed and
approved by The University of Texas at Austin in accordance with its conflict-of-interest policies.

\section{References}


\appendix
\section{Numerical Settings for GENE simulations}
\label{appendix:numerical}

Numerical details of the \textsc{Gene} simulations are described in this appendix.  All simulations were electromagnetic and employed a Landau-Boltzmann collision operator with collision frequencies defined by the experimental conditions.  

Global simulations use 256-384 radial grid points.  The radial domain was $\rho_{tor}=0.898-0.998$ for the narrow pedestal cases and $\rho_{tor}=0.838-0.998$ for the wide pedestal case.  Dirichlet boundary conditions were enforced at the radial boundaries and transition regions were implemented (10 \% on each side) over which gradients are smoothly set to zero and Krook damping smoothly ramps up to set fluctuations to zero at the boundary.  For local simulations, 7 $k_x$ modes were used.

In the parallel $z$ direction, $60-128$ grid points were employed.  In parallel velocity $v_{||}$, 48 grid points were used for local simulations and 60 grid points were used for global simulations.  For the magnetic moment $\mu$ coordinates (i.e. squared perpendicular velocity), 16 grid points were used for local and 24 grid points for global.

\cleardoublepage

\end{document}